\documentclass[acmtog, nonacm]{acmart}
\AtBeginDocument{%
  }

\usepackage{amsmath}    
\usepackage{graphicx}
\usepackage{subcaption}  % for subfigures
\usepackage{multirow}
\usepackage{tabularx}
\usepackage{hyperref}
\newcommand{\hm}[1]{\textcolor{black}{#1}}

\newcolumntype{Y}{>{\centering\arraybackslash}X}
\newcolumntype{C}[1]{>{\centering\arraybackslash}p{#1}}
\usepackage[ruled]{algorithm2e} % For algorithms

\SetAlFnt{\small}
\SetAlCapFnt{\small}
\SetAlCapNameFnt{\small}
\SetAlCapHSkip{0pt}

\usepackage{xcolor}
\begin{document}
\setcopyright{cc}
\setcctype{by-nc-nd}
% \acmJournal{TOG}
% \acmYear{2026} \acmVolume{45} \acmNumber{6} \acmArticle{240}
% \acmMonth{12} \acmDOI{10.1145/3842516}
\title{Length-varying Neural Motion Stitching via Cluster Transition Graph}

\author{Haemin Kim}
\email{spot1223@kaist.ac.kr}
\orcid{0009-0008-1016-4793}
\affiliation{%
  \institution{Korea Advanced Institute of Science and Technology (KAIST)}
  \city{Daejeon}
  \country{Republic of Korea}
}

\author{Junghyun Nam}
\email{ys4990@kaist.ac.kr}
\orcid{0009-0003-6427-2013}
\affiliation{%
  \institution{Korea Advanced Institute of Science and Technology (KAIST)}
  \city{Daejeon}
  \country{Republic of Korea}
}

\author{Seokhyeon Hong}
\email{ghd3079@kaist.ac.kr}
\orcid{0000-0002-8490-5338}
\affiliation{%
  \institution{Korea Advanced Institute of Science and Technology (KAIST)}
  \city{Daejeon}
  \country{Republic of Korea}
}

\author{Vanessa Tan}
\email{vanessa.tan@kaist.ac.kr}
\orcid{0009-0001-8174-6909}
\affiliation{%
  \institution{Korea Advanced Institute of Science and Technology (KAIST)}
  \city{Daejeon}
  \country{Republic of Korea}
}

\author{Junyong Noh}
\email{junyongnoh@kaist.ac.kr}
\orcid{0000-0003-1925-3326}
\affiliation{%
  \institution{Korea Advanced Institute of Science and Technology (KAIST)}
  \city{Daejeon}
  \country{Republic of Korea}
}


\begin{abstract}
Motion stitching aims to create new character animations by seamlessly combining existing motion sequences.
Existing approaches often require manual selection of transition range or assume fixed transition length, restricting the types of motions that can be connected.
To broaden the diversity of motions that can be synthesized, it is essential to generate transitions of varying lengths, allowing the character sufficient time to adapt its pose when the input motions differ significantly.
To this end, we propose a length-varying neural motion stitching method based on a cluster transition graph, which produces naturally connected motion sequences given two distinct input motions.
Our framework consists of three stages: motion clustering, cluster pathfinding, and motion generation.
First, motion clustering maps input motions to discrete clusters.
Next, we identify the corresponding clusters in the cluster transition graph and search for a connecting path.
In this graph, nodes represent motion clusters, and directed edges indicate valid transitions between them.
The resulting path determines both the transition length and a guide sequence that informs motion generation.
Finally, the path and input motions are provided to a Transformer encoder-based motion generator to produce the final transition poses.
Experimental results demonstrate that our method adaptively adjusts the motion length and successfully generates plausible transitions between distinct motions, such as crawling, basketball shooting, and slow locomotion.
We also show that using a graph structure effectively estimates transition durations and produces high-fidelity results compared to methods that assume a fixed transition length, or directly compute the time.
\hm{Code is available at \href{https://haem-k.github.io/nms}{\textbf{\textit{Project Page}}}}
\end{abstract}

\begin{CCSXML}
<ccs2012>
<concept>
<concept_id>10010147.10010371.10010352.10010380</concept_id>
<concept_desc>Computing methodologies~Motion processing</concept_desc>
<concept_significance>500</concept_significance>
</concept>
</ccs2012>
\end{CCSXML}

\ccsdesc[500]{Computing methodologies~Motion processing}

\keywords{Transition generation, Motion stitching, Motion clustering, Motion generation}

\begin{teaserfigure}
  \centering
  \includegraphics[width=\textwidth]{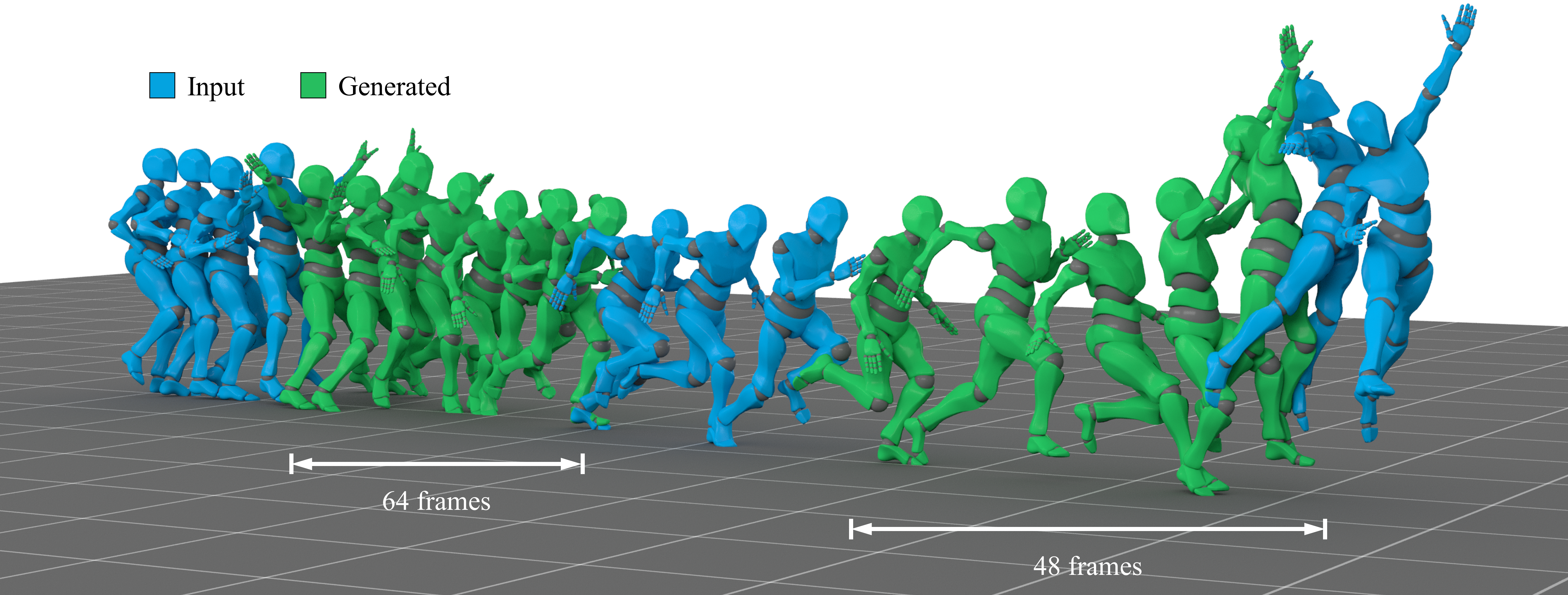}
  \caption{Motion stitching results produced by our method. 
  The results correspond to two test samples. 
  The blue characters indicate the given input motion, while the green characters represent generated transition frames. 
  Depending on the given motion pair, our method generates transitions of varying lengths. 
  In the first transition, the character makes a small jump to close the legs and crouch to match acceleration.
  In the second transition, the character lowers the body to prepare for a big jump in the following volleyball spiking motion.
  The sequence is visualized every 10 frames and the root is translated for clear visualization.}
  \label{fig:teaser}
\end{teaserfigure}

\maketitle
\section{Introduction}

Motion stitching plays a critical role when connecting two different motion sequences in a desired order. 
Through this technique, artists can reuse available motion capture sequences or manually keyframed animations.
A common choice for motion stitching is interpolation-based pose blending, as it is adopted by many animation tools.
Given two input sequences, these methods generally align the second motion to the first motion and then interpolate the poses at the boundary to create intermediate poses.
They produce natural results in cases where the two inputs are sufficiently similar.

Although interpolation-based motion stitching provides a simple approach for connecting motions, it also has notable drawbacks.
For example, the world position and orientation of the second motion are often required to be carefully adjusted to ensure visually pleasing results. 
As an alternative, frame-by-frame pose inspection is performed to find an appropriate transition point.
Even with such manual adjustments, artifacts such as foot sliding or body floating can still occur, especially when the two motions differ significantly.

Kim et al.~\shortcite{kim2023recurrent} suggested a neural network-based method to bypass the process of identifying transition points and support locomotion stitching.
Given two motions that are naively aligned, their method outputs a naturally connected sequence with the same length as the aligned input. 
Unfortunately, their method still assumes two sufficiently similar input motions as it relies on k-nearest \hm{neighbor (KNN)} search to construct input data.
In addition, the lengths of training motion data are fixed, which restricts the types of motions that can be connected.
As a result, the system struggles to handle transitions between two distinct motions such as transitioning from running to lying down, which would require a complex sequence of intermediate motions, including slowing down to a walk, crouching, and gradually lowering the body to match the lying down pose. 

To connect such distinct motions, it is essential to produce length-varying motion sequences.
Prior methods suggested computing transition time from pose differences~\cite{wang2008synthesis}, or training a separate RNN to predict the duration~\cite{guo2022generating}.
However, these approaches often yield inaccurate transition durations due to the lack of world-space cues.
As a result, the generated motion length may not reflect the time required for a character to reach a given pose.
Therefore, instead of explicitly predicting transition time, it is more viable to infer the duration by considering how a character would move to perform the given motions in order.

%%% goal
In this paper, we propose a length-varying neural motion stitching method with a cluster transition graph.
Given two distinct motions arranged for stitching, we focus on producing the shortest transition that varies in length depending on the combination of the two inputs. 
To achieve this, we integrate three core components: motion clustering, cluster pathfinding, and motion generation.
First, we map the two input motions to clusters, and identify them in the cluster transition graph where each node represents a motion cluster.
The cluster corresponding to the first motion becomes the start node, and the one mapped from the second motion becomes the goal node.
We then perform cluster pathfinding by applying a shortest-path algorithm to the graph, using a cost function derived from motion features.
The resulting path of clusters represents the intermediate transition states, which determines the final length of motion.
Using the original input motions and the cluster path, we construct the input for our Transformer encoder-based motion generator \cite{vaswani2017attention}.
The generator outputs a transition sequence connecting the two input motions, along with a destination of the root to align the second input motion.
The final result is obtained by combining the input motions and generated transition frames in order.

% contribution. why graph useful?
A key contribution of our work is the use of a cluster transition graph to guide motion generation.
This graph enables the discovery of intermediate motions, which in turn allows us to estimate transition length and produce high-quality results.
Given a cluster path on the graph, we extract a representative sequence from each cluster using a variant of the Vector Quantized Variational \hm{Autoencoder (VQ-VAE)}~\cite{van2017neural}.
These sequences are then aligned to serve as a guiding signal for the motion generator.
The path determines the total number of frames, and the guide provides per-pose initialization, reducing ambiguity during generation.
As a result, graph-based approach enables automatic motion stitching across a wide range of motion types.

Our contributions are summarized as follows:
\begin{itemize}
    \item We propose a neural motion stitching method that generates variable-length transitions by accounting for structural differences between input motions, thereby extending the range of motions that stitching methods can handle.
    \item The proposed pathfinding approach traverses on the cluster transition graph to implicitly estimate appropriate transition lengths and obtains a guide that disambiguates the motion generation process.
    \item The proposed motion generator effectively synthesizes smooth transitions and also predicts the root destination required to seamlessly align the second motion after transition frames.
\end{itemize}

\section{Related Work}

\subsection{Motion Stitching}
One of the most common techniques for connecting two distinct motions is to interpolate poses at the boundary. 
Owing to their simplicity and efficiency, such approaches have been widely adopted in practical animation systems.
A representative example is Inertialization \hm{Blending (IB)}~\cite{bollo2018inertialization},
which interpolates the displacement between the last pose of the first motion and the first pose of the second motion.
The scaled delta values are then applied to the subsequent poses during the blending interval given by the user.

In addition to obtaining transition poses, determining an appropriate transition length is known to be important in generating naturally connected motion.
Wang et al. \shortcite{wang2008synthesis} highlighted that overly short blend durations result in abrupt changes in velocity, while overly long durations smooth out high-frequency details.
They showed how blend lengths can be computed using geodesic distance or joint velocity.
We similarly emphasize the importance of transition length; however, instead of explicitly calculating the duration in seconds, we estimate the transition length indirectly through graph traversal.

Several methods also explore variable-length transitions instead of directly computing transition time.
Li et al. \shortcite{li2008laziness} introduced elastic L-score based on pose-wise features and minimized the score to insert an adequate number of intermediate frames.
Similarly, Koyama et al. \shortcite{koyama2019precomputed} proposed inserting a "one-hop" sequence of poses between motions, which is chosen based on per-pose features.
Our approach generalizes this idea by inserting multiple motion clips between the input motions using a cluster transition graph, thereby accommodating a wider range of motion types.

In learning-based approaches, Kim et al.~\shortcite{kim2023recurrent} proposed the Recurrent Motion \hm{Refiner (RMR)} that utilizes an RNN to generate smooth transitions.
Their method focuses solely on locomotion data based on KNN-based data pair generation.
In contrast, we aim to incorporate complex movements by building clusters out of a large motion dataset and obtaining variable-length transitions that adapt to diverse input motion pairs.

\subsection{Motion Quantization}
Motion quantization via VQ-VAE has primarily been used to bridge motion with other domains such as text and control signals.
Text-to-motion research adopts VQ-VAE to tokenize motion data and associate them with text.
Using this discrete motion representation, neural networks are trained to generate token sequences either autoregressively~\cite{guo2022tm2t, zhang2023generating, jiang2023motiongpt, lu2025scamo} or through masked modeling~\cite{guo2024momask, pinyoanuntapong2024mmm}.
On the other hand, \citeN{starke2024categorical} used a codebook to align control signals with future motion, and \citeN{li2024walkthedog} proposed a shared codebook for heterogeneous motion datasets with different character morphologies.

Unlike prior work, we use quantization for motion clustering, inspired by \citeN{cho2021motion}.
\citeN{cho2021motion} applied VQ-VAE to cluster motion data into a finite set of groups, thereby reducing the action space in a deep reinforcement learning framework.
Similarly, our purpose of using VQ-VAE is to group short motion segments into a finite number of clusters.
This allows large motion datasets to be abstracted into discrete motion states and represented as a graph without manual labeling.

\subsection{Graph-based Motion Synthesis}
Graph-based motion synthesis has been extensively studied in the field of character animation. 
These approaches typically begin by constructing a graph structure from a given motion dataset.
New motion sequences are then synthesized by traversing the graph, guided by user constraints or cost functions.
%%%
Many existing methods computed pairwise distances between all possible pose combinations in the database to identify natural transition points \cite{10.1145/566654.566605, lee2002interactive, peng2007interactive, heck2007parametric, 10.1145/1276377.1276510, ren2010human}.
In contrast, we represent each cluster of motion blocks as a single node and connect nodes based on whether a valid transition exists between clusters.

Several approaches avoid explicit computation of pose-to-pose distances when constructing motion graphs.
Min et al. \shortcite{min2012motion} manually annotated a few example motions and then identified similar keyframes across the dataset to divide sequences into groups of motion segments.
Likewise, Tao et al. \shortcite{tao2023neural} manually divided the motion dataset and trained a separate mixture-of-experts model for each group.
In contrast to these manually organized datasets, we aim to support a broad spectrum of motions by grouping motion blocks in an unsupervised fashion, using a variant of VQ-VAE as a clustering framework.

\subsection{Motion In-betweening}
In terms of connecting two different frames, motion in-betweening methods can be applied to a motion stitching task. 
Early works such as Harvey et al. \shortcite{harvey2020robust} utilized an RNN to synthesize in-between frames. 
Their model encoded the first 10 context frames, a target keyframe, and pose offsets to predict in-between poses. 
This approach was further extended by combining a Variational Autoencoder \cite{tang2022real} and style embeddings \cite{Tang_2023}.

% Dense context frames: Transformer-based
To mitigate error accumulation inherent in autoregressive models, several studies have employed Transformer-based architectures, predicting the entire motion sequence at once. 
\citeN{Kim_2022} enhanced controllability by including an anchor pose and semantic embeddings. \citeN{qin2022motion} and \citeN{hong2024long} proposed a two-stage Transformer framework, where the first network produces coarse pose estimates and the second refines them into detailed motion. 
\citeN{oreshkin2023motion} presented the Delta-interpolator, which predicts delta motion vectors to be applied to interpolated poses.
\citeN{akhoundi2025silk} investigated effective data representation for the motion in-betweening task.
They showed that selecting an appropriate motion format can lead to comparable performance to recent methods, even with a simple Transformer encoder architecture.
Our motion generator shares similarities with the model proposed by \citeN{akhoundi2025silk}. 
Instead of setting all in-between frames as zero, we integrate guide information derived from the result of our pathfinding module. 
This enables our model to produce transitions across a diverse set of motion types while maintaining consistent quality.

% Sparse frames: Transformer-based, Diffusion-based model 
In contrast to methods that utilize dense context frames, several approaches adopt a minimal in-betweening setup with only a single context frame and a target frame.
\citeN{he2022nemf} proposed neural motion fields that generate poses conditioned on time, supporting both sparse and dense context frames.
\citeN{starke2023motion} and \citeN{chu2024real} generated poses autoregressively using a mixture-of-experts architecture, and \citeN{agrawal2024skel} employed skeletal Transformer layers.
In parallel, diffusion-based methods have been widely explored for controllable motion synthesis under joint constraints~\cite{tevethuman, karunratanakul2023guided, xie2023omnicontrol, dai2024motionlcm, bae2025less, pinyoanuntapong2025maskcontrol}.
\citeN{cohan2024flexible} incorporated a masking strategy to further specialize in the motion in-betweening task.

To obtain high-quality outputs from motion in-betweening methods, it is important to provide a plausible target pose in the world frame, along with precise timing that allows the character to reach the desired state.
Goel et al.~\shortcite{goel2025generative} noted the practical difficulty for users in supplying intermediate keyframe inputs at precise timings. 
To address this, they proposed learning time warp parameters jointly with pose predictions, automatically adjusting the keyframe timing. 
Our approach automatically determines the total motion length and the transformation of the second motion, without requiring users to specify explicit values.

\begin{figure*}
    \centering
    \includegraphics[width=0.95\textwidth]{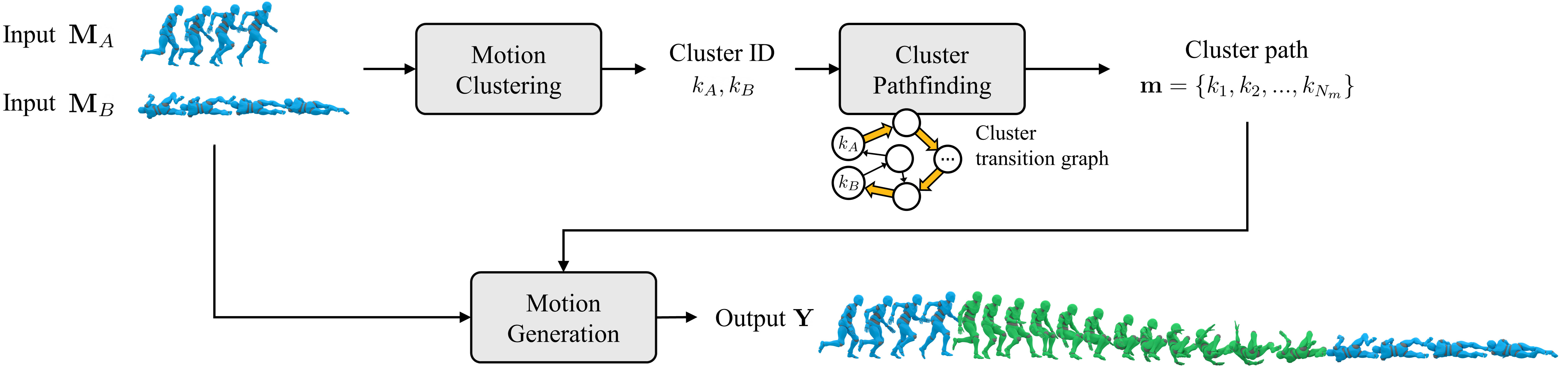}
    \caption{System overview. Input motions $\mathbf{M}_A$ and $\mathbf{M}_B$ are mapped to cluster ID $k_A$ and $k_B$ through motion clustering.
    During cluster pathfinding, we run a shortest-path algorithm on the cluster transition graph, considering $k_A$ as the start node and $k_B$ as the goal node. 
    Once we obtain the cluster path $\mathbf{m}$, we utilize the path and the original input motions to generate final connected motion $\mathbf{Y}$.}
    \label{fig:overview}
\end{figure*}

\section{Overview}\label{sec:method}
Our framework connects two input motions $\mathbf{M}_A$ and $\mathbf{M}_B$ into an output motion $\mathbf{Y}$ through three main stages.
We assume that each input is a raw motion consisting of $T_l+1$ frames, where the additional frame is used to compute displacement over $T_l$ frames.
Both inputs are defined in canonical space, approximately aligned with the identity transformation in the world frame.
Specifically, the first pose of the sequence faces the forward \hm{direction (z-axis)} of the world frame and the x and z components of its root position starts at the origin on the ground \hm{plane (xz-plane)}.
As shown in Figure~\ref{fig:overview}, we first perform motion clustering to assign each input motion to a cluster, mapping $\mathbf{M}_A$ to cluster ID $k_A$ and $\mathbf{M}_B$ to $k_B$.
A cluster consists of \textit{motion blocks}, short motion segments of $T_l$ frames, which serve as the basic building blocks of motion throughout our method.
In the cluster pathfinding stage, we employ a cluster transition graph where $k_A$ and $k_B$ become the start and goal nodes, respectively.
A shortest-path algorithm then finds a cluster path $\mathbf{m}=\{{k_1}, {k_2}, \cdots, {k_{N_m}}\}\in\mathbb{R}^{N_m}$, where $k_1=k_A$ and $k_{N_m}=k_B$, and determines the number of motion blocks $N_m$ that constitute the entire sequence.
During motion generation, we aim to generate only the transition frames and then combine them with the original input motions.
To this end, we use $\mathbf{M}_A$, $\mathbf{M}_B$, and $\mathbf{m}$ as input to the motion generator, which produces the final output motion $\mathbf{Y}$ with length $T={T_l\cdot{N_m}}$.
With all three stages combined, we refer to the overall framework as Neural Motion \hm{Stitching (NMS)}.

\begin{figure}[b]
    \centering
    \includegraphics[width=0.42\textwidth]{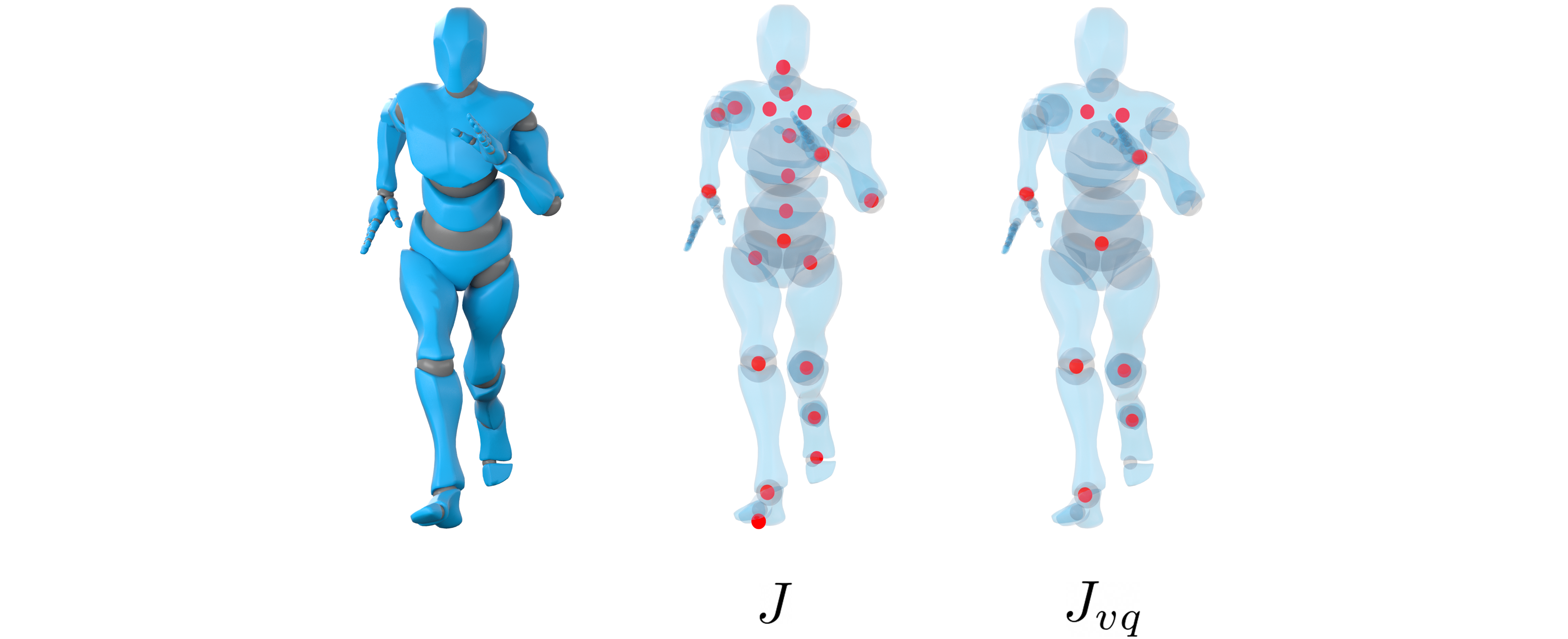}
    \caption{Joints used in the method. Our basic definition for a skeleton includes $J=22$ joints with hip joints as the root. 
    For CVQ-VAE, we use $J_{vq}=9$ and for adversarial training, we also use the head joint which makes $J_{vq}+1=10$.}
    \label{fig:joints}
\end{figure}

Throughout our method, we fix a motion block length based on 60 FPS motion data.
The minimum transition length is fixed to 0.3 seconds following Kovar et al.~\shortcite{10.1145/566654.566605}, and accordingly the motion block length becomes $T_l=16$ \hm{frames ($\approx 0.3$s)}, on the premise that multiple motion blocks are combined to form long transitions between distinct motions.
For the skeleton, we define the hip joint as the root of the character, and use $J=22$ joints for the full skeleton motion generation and $J_{vq}=9$ for VQ-VAE training as shown in Figure~\ref{fig:joints}.
All orientation values used for network training are represented in 6D format~\cite{zhou2019continuity}.

\section{Motion Clustering}

\begin{figure}
    \centering
    \includegraphics[width=0.47\textwidth]{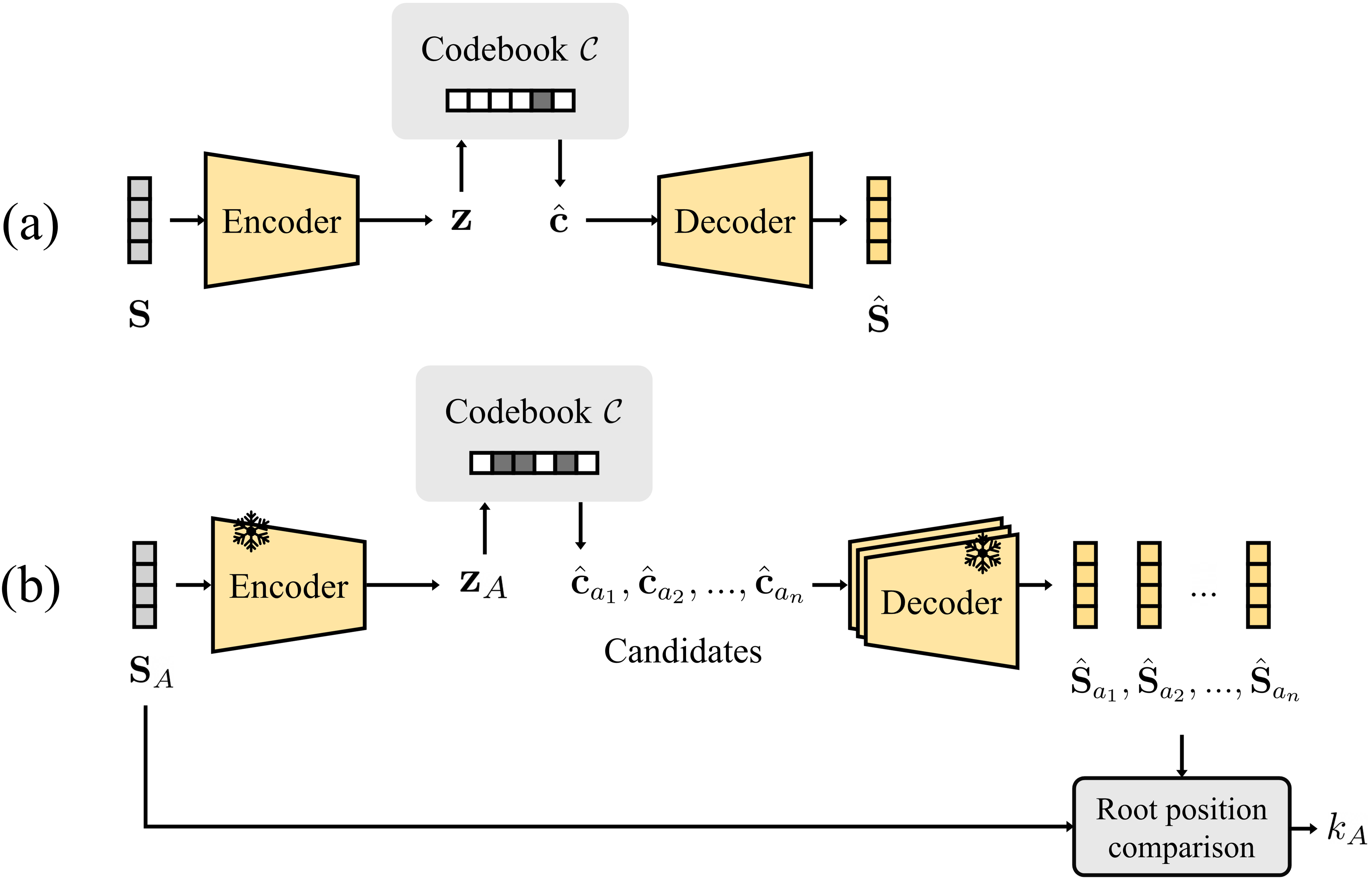}
    \caption{Motion clustering. Figure (a) shows CVQ-VAE training process for cluster construction. Figure (b) shows the cluster assignment process after training.}
    \label{fig:clustering}
\end{figure}

% clustering
The goal of motion clustering is to categorize given motions $\mathbf{M}_A$ and $\mathbf{M}_B$ into groups.
We first construct motion clusters and then establish a method to map new motion blocks to one of the clusters.
Inspired by \citeN{cho2021motion}, we train Clustering \hm{VQ-VAE (CVQ-VAE)} \cite{zheng2023online} to construct motion clusters in an unsupervised manner.
The CVQ-VAE training process is shown in \hm{Figure~\ref{fig:clustering} (a)}.
The input motion block $\mathbf{S}=
[\mathbf{s}_1\,\, \mathbf{s}_2\,\, \dots\,\, \mathbf{s}_{T_l}]\in\mathbb{R}^{{T_l}\times{d_{s}}}$ is composed of $T_l$ pose vectors $\mathbf{s}_t\in\mathbb{R}^{d_s}$, where $d_s$ is the dimension of a pose vector.
The encoder converts $\mathbf{S}$ into a vector $\mathbf{z}\in\mathbb{R}^{d_c}$, and it is mapped to the nearest code $\hat{\mathbf{c}}\in\mathbb{R}^{d_c}$ in the codebook $\mathcal{C}=\{\mathbf{c}_1, \mathbf{c}_2, \dots, \mathbf{c}_{N_c}\}\subset\mathbb{R}^{d_c}$ \hm{in terms of cosine similarity}, where $N_c$ is the codebook size and $d_c$ is the codebook dimension. 
Each code represents a unique motion cluster, and we refer to these codes as \textit{cluster codes}.
The selected code $\hat{\mathbf{c}}$ is then decoded into a motion block $\hat{\mathbf{S}}=[\hat{\mathbf{s}}_1\,\, \hat{\mathbf{s}}_2\,\, \dots\,\, \hat{\mathbf{s}}_{T_l}]\in\mathbb{R}^{{T_l}\times{d_{s}}}$.
We define this decoded sequence as a \textit{cluster clip}, which serves as the representative sequence for the corresponding cluster.

As illustrated in \hm{Figure~\ref{fig:clustering} (b)}, the cluster ID for a new motion is assigned using the trained CVQ-VAE.
We adopt a candidate-based approach with root position comparison to reduce the risk of outliers and prevent false edge connections during graph construction.
First, we convert the representation of the given motion $\mathbf{M}_A$ into $\mathbf{S}_A$ and encode it to $\mathbf{z}_A$.
The top $n$ candidate codes $\hat{\mathbf{c}}_{a_1}, \hat{\mathbf{c}}_{a_2}, \ldots, \hat{\mathbf{c}}_{a_n}$ are retrieved according to their cosine similarity to $\mathbf{z}_A$, where $a_1, a_2, \ldots, a_n$ are the candidate cluster IDs.
We decode all candidate codes and acquire corresponding cluster clips, $\hat{\mathbf{S}}_{a_1}, \hat{\mathbf{S}}_{a_2}, \ldots, \hat{\mathbf{S}}_{a_n}$.
The final cluster ID $k_A$ is selected from the candidate cluster IDs by comparing the root position between the cluster clips and $\mathbf{S}_A$.
We set the number of candidate to $n=3$ in all of our experiments.
The assignment process is repeated with $\mathbf{M}_B$ to obtain its cluster ID, $k_B$.
For more details on cluster assignment, please refer to Section~A of the supplementary material.

The number of clusters $N_k$ equals the number of codes selected during cluster assignment.
It is heavily affected by the codebook size $N_c$, as we adopt the online reinitialization technique suggested by \citeN{zheng2023online} during training to maximize codebook usage and improve reconstruction quality.
In our implementation, we use $N_c=2048$, achieving 100\% codebook utilization.
Thus, the total number of clusters is $N_k=N_c=2048$.

% ------------------------------------------------------------
\subsection{Cluster Construction}

\subsubsection{Data preprocessing}
We prepare a training dataset for CVQ-VAE using $J_{vq}$ key joints that play a critical role in visually distinguishing motions.
We segment the motion capture sequences into clips of $T_l+1$ frames, sliding the window forward by $T_l/2$.
Instead of using the segmented clips directly, we curate the important features and convert them to motion blocks $\mathbf{S}$.
Each pose vector $\mathbf{s}_t$ at frame $t$ is defined as follows:
\begin{equation} 
\mathbf{s}_t=\{r^{h}_t, 
\mathbf{r}^{q}_t, 
\mathbf{r}^{\Delta{p}}_t, 
\mathbf{r}^{\Delta{q}}_t, 
\mathbf{j}^{prel}_t, 
\mathbf{j}^{\Delta{p}}_t\}
\in\mathbb{R}^{d_{s}}
\nonumber
\end{equation}
\noindent
\sloppy where 
$r^{h}_t$ is the root height, 
$\mathbf{r}^{q}_t\in\mathbb{R}^6$ is the root orientation,
$\mathbf{r}^{\Delta{p}}_t\in\mathbb{R}^3$ is the root translational displacement,
$\mathbf{r}^{\Delta{q}}_t\in\mathbb{R}^6$ is the root angular displacement,
$\mathbf{j}^{prel}_t\in\mathbb{R}^{3J_{vq}}$ is the root-relative joint position,
and $\mathbf{j}^{\Delta{p}}_t\in\mathbb{R}^{3J_{vq}}$ is the joint translational displacement.
As shown in Figure~\ref{fig:joints}, we pick a subset of joints by following Cho et al. \shortcite{cho2021motion} and add knee joints to help distinguish actions such as crawling and rolling on the ground, resulting in $J_{vq}=9$.
We compute translational displacement by subtracting the joint's world position at frame $t$ from its position at frame $t+1$, and angular displacement by taking the relative rotation from frame $t$ to $t+1$.

\subsubsection{Network architecture}
We utilize a simple autoencoder architecture based on convolutional layers.
The encoder compresses the time dimension of the input $\mathbf{S}$ to $T_l/4$ through nine 1D convolutional layers with residual connections.
The compressed tensor is flattened into a one-dimensional vector and passed through two linear layers, producing a vector $\mathbf{z}$.
We use cosine distance to find the closest vector $\hat{\mathbf{c}}$ of dimension $d_c=64$ in the codebook $\mathcal{C}$.
The selected vector $\hat{\mathbf{c}}$ is passed through two linear layers again, and fed into a decoder with an architecture symmetric to the encoder, reconstructing it into a sequence $\hat{\mathbf{S}}$.

\subsubsection{Loss function}
Our loss function for training CVQ-VAE is as follows:
\begin{align}
\mathcal{L}_{cluster} 
&= \lambda_{recon}\mathcal{L}_{recon} \, + \, \lambda_{pos}\mathcal{L}_{pos} \notag \\
&\quad + \, \lambda_{vq}\mathcal{L}_{vq} \, + \, \lambda_{contras}\mathcal{L}_{contras},
\notag
\end{align}
\noindent
where $\lambda_{recon}$, $\lambda_{pos}$, $\lambda_{vq}$, and $\lambda_{contras}$ controls the scale of each loss and set to 0.01, 1.5, 1.0, 1.0, respectively.

% reconstruction loss
The reconstruction loss $\mathcal{L}_{recon}$ computes the mean squared error between poses in the input $\mathbf{S}$ and those in the output $\hat{\mathbf{S}}$:
\[
\mathcal{L}_{recon} = 
\frac{1}{T_l} \sum_{t=1}^{T_l}\|\mathbf{s}_{t} - \hat{\mathbf{s}}_{t}\|_2^2,
\]
\noindent
where each features are normalized with mean and standard deviation of each channel.

% position loss
Additionally, we compute the position loss $\mathcal{L}_{pos}$ in the original data distribution to capture movement in the world frame:
\[
\mathcal{L}_{pos} = 
\frac{1}{T_l} \sum_{t=1}^{T_l}\|\mathbf{T}_t\mathbf{j}^{prel}_{t} - \hat{\mathbf{T}}_t\hat{\mathbf{j}}^{prel}_{t}\|_2^2,
\]
\noindent
where $\mathbf{T}_t$ is the root transformation reconstructed from $r^{h}_t$, $\mathbf{r}^{\Delta{p}}_t$, and $\mathbf{r}^{q}_t$.
We multiply $\mathbf{T}_t$ to $\mathbf{j}^{prel}_t$ to obtain the joint's world position, and compute the mean squared error between the world joint positions in the input $\mathbf{S}$ and the output $\hat{\mathbf{S}}$.

% VQ loss 
The vector quantization loss $\mathcal{L}_{vq}$ encourages encoder outputs and codebook codes to move toward each other~\cite{van2017neural}:
\[
\mathcal{L}_{vq} = 
\|sg[\mathbf{z}]-\hat{\mathbf{c}}\|_2^2
+ \beta\|\mathbf{z}-sg[\hat{\mathbf{c}}]\|_2^2,
\]
where $sg[\cdot]$ denotes the stop-gradient operator and $\beta$ is a weighting factor set to $0.25$ following \citeN{van2017neural}. 

We also adopt the contrastive loss $\mathcal{L}_{contras}$ to promote code sparsity and distinct clusters~\cite{zheng2023online}:
% \[
% \mathcal{L}_{contras}=-\log \frac{e^{sim(e_{k}, \hat{z}_{i}^{+})/\tau}}{\sum_{i=1}^{N}e^{sim(e_{k}, \hat{z}_{i}^{-})/\tau}}
% \]
\[
\mathcal{L}_{contras}=-\frac{1}{N_c}\sum_{i=1}^{N_c}\log \frac{e^{sim(\mathbf{c}_{i}, \mathbf{z}_{i}^{+})/\tau}}{e^{sim(\mathbf{c}_{i}, \mathbf{z}_{i}^{+})/\tau} + \sum_{j=1}^{N_{neg}}e^{sim(\mathbf{c}_{i}, \mathbf{z}_{j}^{-})/\tau}}
\]
where $sim(\cdot)$ denotes cosine similarity, $N_{neg}$ is the number of negative samples in a training batch, and $\tau$ is the temperature parameter set to $0.07$.
For each cluster code $c_{i}$, we treat the closest latent vector $\mathbf{z}_{i}^{+}$ in the batch as the positive sample and the bottom 50\% of the latent vectors with the lowest similarity as negative samples.

\section{Cluster Pathfinding}

\subsection{Graph Construction}
\begin{figure}
    \centering
    \includegraphics[width=0.42\textwidth]{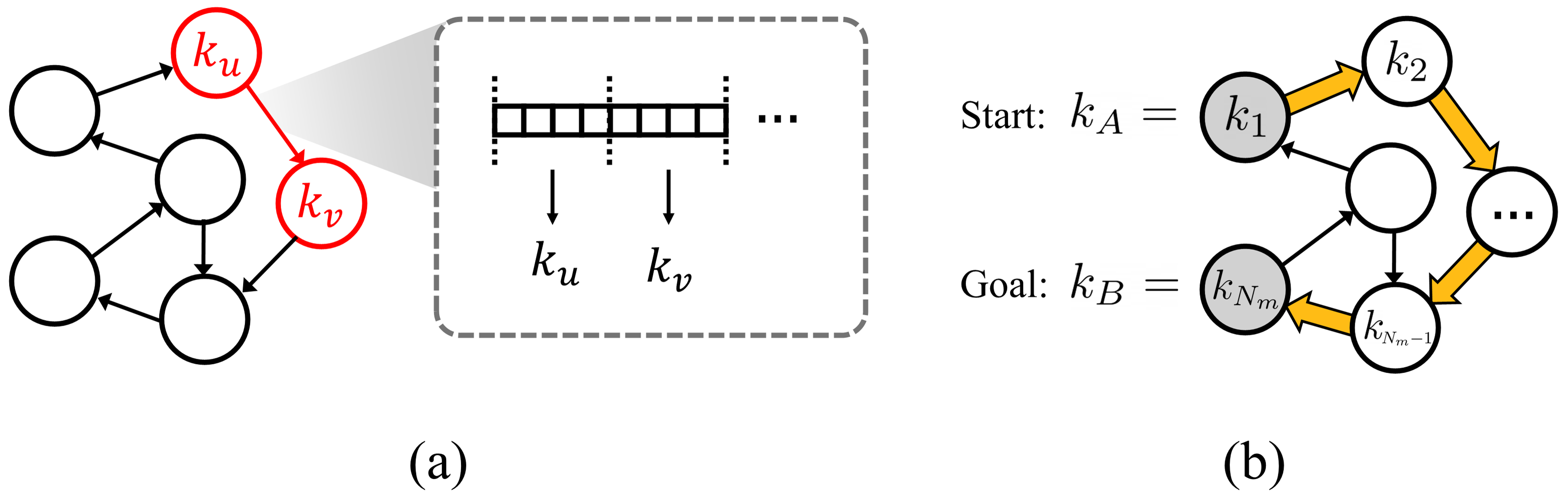}
    \caption{Cluster pathfinding. Figure (a) visualizes how edge is defined in the cluster transition graph. Figure (b) shows the result of pathfinding algorithm. The yellow arrows indicate selected edges for the shortest path.}
    \label{fig:path}
\end{figure}

The objective of cluster pathfinding is to acquire a cluster path $\mathbf{m}$ that specifies cluster clips to be inserted between the two input motions $\mathbf{M}_A$ and $\mathbf{M}_B$.
To this end, we first construct the cluster transition graph, a directed graph in which each node represents a cluster and each edge denotes a valid transition between the clusters.
We build this graph by sliding a window across the motion capture sequences, assigning a cluster to each window, and defining edges between consecutive clusters.
For example, as described in \hm{Figure~\ref{fig:path} (a)}, we assign a cluster $k_u$ to a block in the motion capture sequence, slide the window forward without overlapping, and assign a cluster $k_v$ to the next block.
Because these two blocks are adjacent in the motion capture sequence, we define a directed edge from cluster $k_u$ to $k_v$.
By connecting these two clusters, we assume that there exists a plausible transition from $k_u$ to $k_v$.
Repeating this process across all training sequences yields a directed graph with $N_k$ nodes.

\subsection{Graph Traversal}\label{traversal}
We apply the A* search algorithm to the graph and find the shortest cluster path $\mathbf{m}$ consisting of $N_m$ nodes, which starts at node $k_A$ and ends at node $k_B$ as illustrated in \hm{Figure~\ref{fig:path} (b)}.
This process finds $N_m-2$ nodes that will be inserted between the two inputs, which in turn determine the length of the final motion.
Because edges in the graph are constructed only between clusters with valid transitions, any path corresponds to a sequence of natural motion blocks.
Therefore, the shortest path is the minimal sequence of such blocks connecting the start and goal nodes, instead of the shortest geometric distance in joint space.
We choose this path to avoid unnecessary intermediate motion states.
For instance, given two walking motions, a natural expectation is a walking-to-walking transition instead of passing through states such as sitting or running.
This formulation ensures that the generated transitions remain consistent with the input motions.

The A* algorithm selects a path that minimizes:
\[
f(k_u)=g(k_u)+h(k_u),
\]
\noindent
where $k_u$ is the next node on the path, $g(k_u)$ is the accumulated cost along the edges from the start node $k_A$ to $k_u$, and $h(k_u)$ is a heuristic function that estimates the cost from $k_u$ to the goal node, $k_B$.

To compute $g(k_u)$, we define a cost function for each edge that accounts for the distance between the connected nodes in the latent space and the pose difference in motion space.
For each edge $(k_u, k_v)$, the cost $d$ is defined as:
\begin{equation} 
\label{eqn:cost}
d(k_u, k_v) = \lambda_{code}d_{code}(\mathbf{c}_{k_u}, \mathbf{c}_{k_v}) + \lambda_{bnd}d_{bnd}(\hat{\mathbf{s}}_{k_u, T_l}, \hat{\mathbf{s}}_{k_v, 1}),
\end{equation}
\noindent
where $\mathbf{c}_{k_u}$ and $\mathbf{c}_{k_v}$ are the respective cluster codes, 
$\hat{\mathbf{S}}_{k_u}=
[\hat{\mathbf{s}}_{k_u, 1}\,\, \hat{\mathbf{s}}_{k_u, 2}\,\, \dots\,\, \hat{\mathbf{s}}_{k_u, T_l}]$ 
and 
$\hat{\mathbf{S}}_{k_v}=
[\hat{\mathbf{s}}_{k_v, 1}\,\, \hat{\mathbf{s}}_{k_v, 2}\,\, \dots\,\, \hat{\mathbf{s}}_{k_v, T_l}]$ 
are the cluster clips with $T_l$ frames.
The first term with $d_{code}$ captures dissimilarity between the two clusters in the latent space by measuring the cosine distance between the cluster codes.
The second term with $d_{bnd}$ evaluates the cost of transitioning at the boundary, measuring the root difference between the last pose of $\hat{\mathbf{S}}_{k_u}$ and the first pose of $\hat{\mathbf{S}}_{k_v}$.
The weights $\lambda_{code}$ and $\lambda_{bnd}$ control the relative importance of these terms and we set $\lambda_{code}=0.2$ and $\lambda_{bnd}=1.0$ in our implementation.

The heuristic $h(k_u)$ is defined in the same manner, but compares $k_u$ and the goal node $k_B$ as follows:
\begin{equation}
\label{eqn:heu}
    h(k_u) = \lambda_{code}d_{code}(\mathbf{c}_{k_u}, \mathbf{c}_{k_B}) + \lambda_{bnd}d_{bnd}(\hat{\mathbf{s}}_{k_u, T_l}, \hat{\mathbf{s}}_{k_B, 1}),
\end{equation}
\noindent
where $\mathbf{c}_{k_B}$ is the cluster code of the goal node, and $\hat{\mathbf{S}}_{k_B}=[\hat{\mathbf{s}}_{k_B, 1}\,\, \hat{\mathbf{s}}_{k_B, 2}\,\, \dots\,\, \hat{\mathbf{s}}_{k_B, T_l}]$ is its cluster clip.
The $\lambda_{code}$ and $\lambda_{bnd}$ are the weights for each term and take the same values as in Equation \ref{eqn:cost}.

Inspired by Lee et al.~\shortcite{lee2002interactive}, we define $d_{bnd}$ using the root features from the cluster clips.
Given two boundary poses 
$\hat{\mathbf{s}}_{k_u, T_l}$
and 
$\hat{\mathbf{s}}_{k_v, 1}$, we measure differences in root height, root orientation, root translational displacement, and root angular displacement:
{
\setlength{\jot}{7pt} % increase line space
\begin{align} \label{eqn:bnd}
    d_{bnd}(\hat{\mathbf{s}}_{k_u, T_l}, \hat{\mathbf{s}}_{k_v, 1}) 
    &=\lambda_h\|r_{k_v, 1}^h - r_{k_u, T_l}^h\|_2 
    \notag \\ 
    &\quad + \lambda_{q}\|\log(\mathbf{R}^{q}_{k_v, 1}(\mathbf{R}^{q}_{k_u, T_l})^{-1})\|_2
    \notag \\
    &\quad + \lambda_{\Delta{p}}\|\mathbf{r}_{k_v, 1}^{\Delta{p}} - \mathbf{r}_{k_u, T_l}^{\Delta{p}}\|_2
    \notag \\
    &\quad + \lambda_{\Delta{q}}\|\log(\mathbf{R}^{\Delta{q}}_{k_v, 1}(\mathbf{R}^{\Delta{q}}_{k_u, T_l})^{-1})\|_2.
\end{align}
}
\noindent
Here, $\mathbf{R}^{q}$ is the root orientation matrix converted from $\mathbf{r}^{q}$, and $\mathbf{R}^{\Delta{q}}$ is the root angular displacement matrix converted from $\mathbf{r}^{\Delta{q}}$.
We only use the root features in $d_{bnd}$, because the root motion has the most significant impact on achieving natural results.
For the weights, we set $\lambda_h=1.0$, 
$\lambda_{q}=1.0$, 
$\lambda_{\Delta{p}}=0.5$, 
and $\lambda_{\Delta{q}}=0.5$ in our implementation.
\section{Motion Generation}
\begin{figure}[b]
    \centering
    \includegraphics[width=0.42\textwidth]{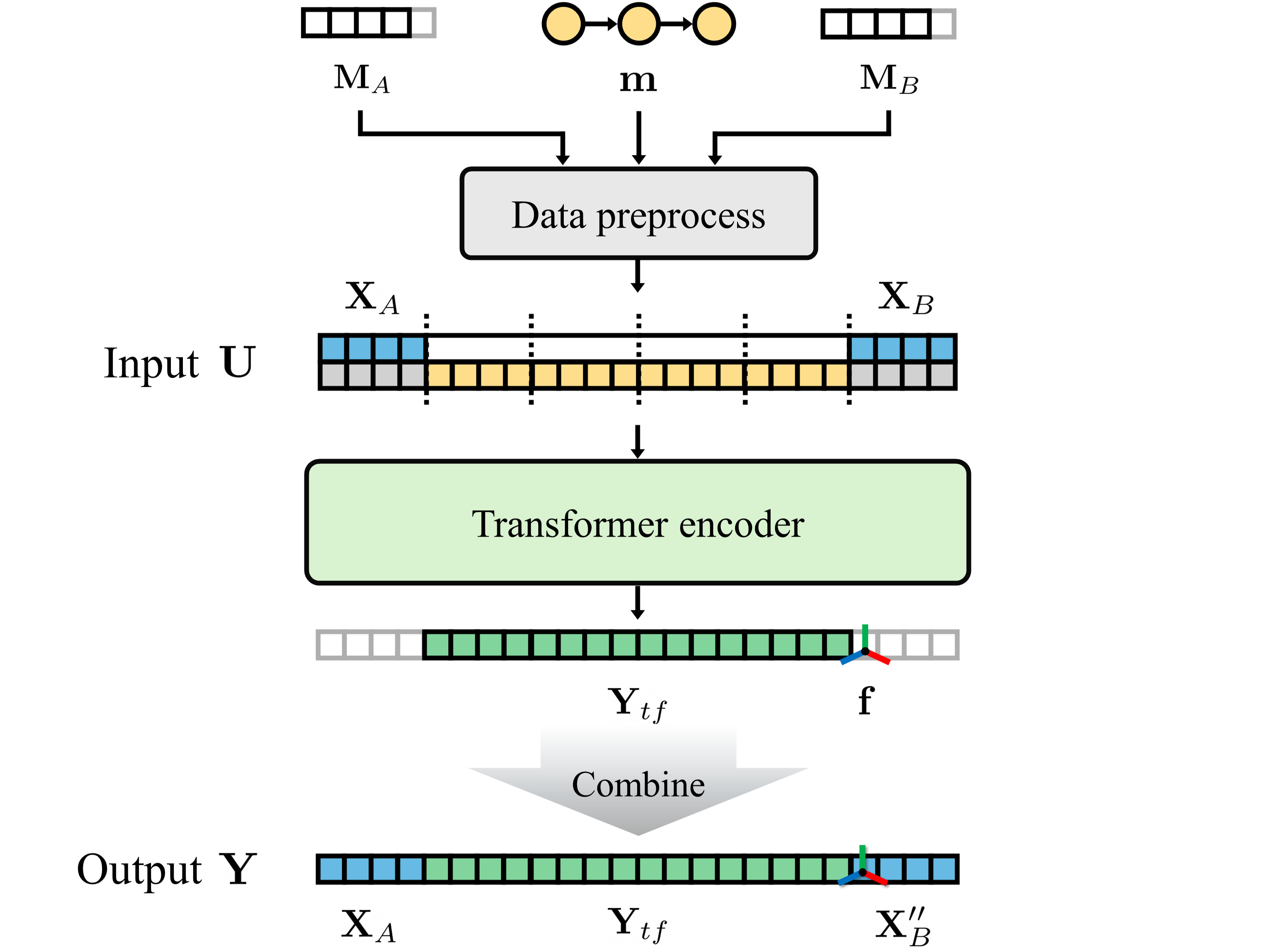}
    \caption{Motion generation. The input motions $\mathbf{M}_A$ and $\mathbf{M}_B$ and the cluster path $\mathbf{m}$ are processed to construct the network input $\mathbf{U}$. The preprocessed sequence $\mathbf{U}$ is then provided to the Transformer encoder, and outputs transition poses $\mathbf{Y}_{tf}$ and the root destination vector $\mathbf{f}$. Lastly, the original input motions and the output are combined to produce the connected result $\mathbf{Y}$.}
    \label{fig:generation}
\end{figure}
% motion generation
Motion generation produces the final connected motion based on the two inputs and the cluster path.
As shown in Figure~\ref{fig:generation}, the inputs $\mathbf{M}_A$, $\mathbf{M}_B$, and $\mathbf{m}$ are preprocessed to form the input $\mathbf{U}\in\mathbb{R}^{T\times (d_x+d_s)}$, where $T$ is the number of frames in the final sequence and $d_x+d_s$ is the dimension of each frame vector.
The sequence $\mathbf{U}$ includes $\mathbf{X}_A\in\mathbb{R}^{T_l\times d_x}$ and $\mathbf{X}_B\in\mathbb{R}^{T_l\times d_x}$, where each input motion $\mathbf{M}_A$ and $\mathbf{M}_B$ is represented as $T_l$ pose vectors $\mathbf{x}\in\mathbb{R}^{d_x}$.
These pose vectors encompass the orientations of all $J$ joints, resulting in a total dimensionality of ${d_x}$ per pose.
A Transformer encoder receives $\mathbf{U}$ and outputs a sequence of the same length, from which we extract only the transition frames $\mathbf{Y}_{tf}$ in the range $T_l+1 \leq t \leq T-T_l$.
We also predict the root destination vector $\mathbf{f}=\{r_{T_{dest}}^h, \mathbf{r}_{T_{dest}}^q, \mathbf{r}_{T_{dest}}^{\Delta{p}}, \mathbf{r}_{T_{dest}}^{\Delta{q}}\}\in\mathbb{R}^{16}$, where $T_{dest}=T-T_l+1$.
To assemble the final motion $\mathbf{Y}\in\mathbb{R}^{T\times {d_x}}$, we transform and concatenate $\mathbf{X}_A$ and $\mathbf{X}_B$ with $\mathbf{Y}_{tf}$.
We convert $\mathbf{f}$ to a transformation matrix and match the first root of $\mathbf{X}_B$ to this transformation to obtain $\mathbf{X}_B''$.
Finally, we concatenate $\mathbf{X}_A$, $\mathbf{Y}_{tf}$, and $\mathbf{X}_B''$ to obtain the connected motion $\mathbf{Y}$.

%%%%%%%%%%%%%%%%%%%%%%%%%%%%%%%%%%%%%%%%%%%%%%%%%%%%%%%%%%%%%%%%
\subsection{Data Preprocessing}
\subsubsection{Data format}
To generate the final motion that includes all $J$ joints and their orientations, we incorporate a data format different from that used in the motion clustering stage.
We define a motion $\mathbf{X}$ as a sequence of pose vectors $\mathbf{x}_t$, where $t$ indicates the frame index.
Then, each pose vector is defined as follows:
\begin{equation} 
\mathbf{x}_t=\{r^{h}_t, 
\mathbf{r}^{q}_t, 
\mathbf{r}^{\Delta{p}}_t, 
\mathbf{r}^{\Delta{q}}_t, 
\mathbf{j}^{qrel}_t\}
\in\mathbb{R}^{d_{x}}
\nonumber
\end{equation}
\noindent
where $\mathbf{j}^{qrel}_t\in\mathbb{R}^{6J}$ is a root-relative joint orientation following \citeN{akhoundi2025silk}.
We use this full joint representation for $\mathbf{X}_{A}$, $\mathbf{X}_{B}$, $\mathbf{Y}_{tf}$ and final output $\mathbf{Y}$.

\subsubsection{Input processing} \label{input}
\begin{figure}
    \centering
    \includegraphics[width=0.45\textwidth]{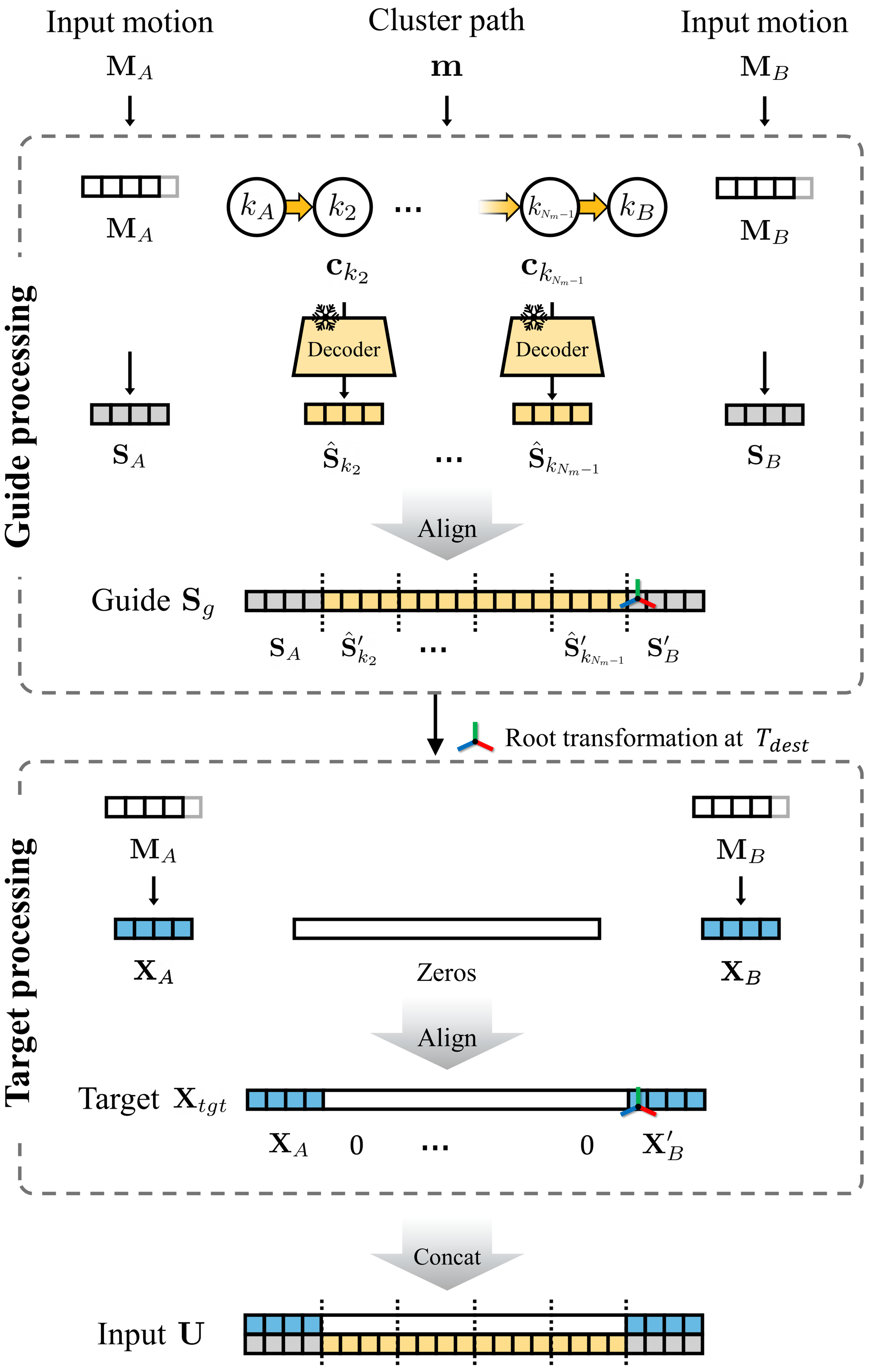}
    \caption{Input processing for motion generation. The first section describes guide processing, and the second shows target processing. 
    The arrow connecting the first and second sections indicates that the root transformation at frame $T_{dest}$ in the guide is copied to translate and rotate $\mathbf{X}_B$.}
    \label{fig:preprocess}
\end{figure}

% data preprocessing
As shown in Figure~\ref{fig:preprocess}, we utilize the raw input motions $\mathbf{M}_A$ and $\mathbf{M}_B$, and a cluster path $\mathbf{m}$ to form the network input $\mathbf{U}$.
To acquire $\mathbf{U}$, we prepare two sequences: a \textit{guide} $\mathbf{S}_g$ and a \textit{target} $\mathbf{X}_{tgt}$.
The guide $\mathbf{S}_g$ supplies the first draft of intermediate poses to the motion generator with information about $J_{vq}$ joints.
It is obtained by combining cluster clips corresponding to each node in the path.
Specifically, \hm{as shown in the "Guide processing" section of Figure~\ref{fig:preprocess}}, the codes $\mathbf{c}_{k_2}, \dots, \mathbf{c}_{k_{N_m-1}}$ are decoded into $N_m-2$ cluster clips, $\hat{\mathbf{S}}_{k_2}, \dots, \hat{\mathbf{S}}_{k_{N_m-1}}$.
We also represent $\mathbf{M}_A$ and $\mathbf{M}_B$ as $\mathbf{S}_A$ and $\mathbf{S}_B$, respectively, matching the representation used for cluster clips.
Then, we align each block to the last root of the previous block, and concatenate them into a sequence in the order: $\mathbf{S}_A$, $\hat{\mathbf{S}}_{k_2}', \dots, \hat{\mathbf{S}}_{k_{N_m-1}}'$, $\mathbf{S}_B'$.
Here, we define motion alignment as translating a motion on the ground plane and rotating it around the up-axis to closely match the given transformation.
The resulting sequence forms the guide $\mathbf{S}_g$, whose length determines the final sequence length $T$.
The accumulated root transformation at frame $T_{dest}$ acts as the initial estimate of the first root transformation of $\mathbf{X}_B$ in the final connected motion.

The target $\mathbf{X}_{tgt}$ provides high-resolution motion data in the target skeleton with $J$ joints, which improves the training convergence of the motion generator.
\hm{The construction of the target is illustrated in the "Target processing" section of Figure~\ref{fig:preprocess}.}
To construct $\mathbf{X}_{tgt}$, we convert $\mathbf{M}_A$ and $\mathbf{M}_B$ into $\mathbf{X}_A$ and $\mathbf{X}_B$, respectively, while keeping $\mathbf{X}_A$ intact and setting the following transition frames to zero.
The root transformation at $T_{dest}$ is then copied from the guide $\mathbf{S}_g$, to align $\mathbf{X}_B$ to this transformation to obtain $\mathbf{X}_B'$.
The aligned sequence is then concatenated after the zeroed transition frames to complete the target.
As a final step, $\mathbf{X}_{tgt}$ and $\mathbf{S}_g$ are concatenated along the feature dimension to produce the network input $\mathbf{U}=[\mathbf{X}_{tgt}\,\,\mathbf{S}_{g}]$.
For more details on input processing, please refer to Section~B of the supplementary material.

\subsubsection{Training dataset}\label{sec:mg_dataset}
For training, we generate variable-length data pairs from motion capture sequences.
Each motion capture sequence is segmented to clips of $m$ motion blocks, which become our ground truth data for the motion generator.
In our implementation, we obtained varying lengths of motions by using $3 \leq m \leq 8$ and applying a sliding window with stride $3\cdot T_l/2$.
To create the paired input, we directly utilize ground truth instead of employing cluster pathfinding.
We split the ground truth into blocks, assign each block to a cluster, and then compose the guide and target sequences as described in Section~\ref{input}.
This approach enables the network to learn motion generation of varying lengths in a supervised manner.

%%%%%%%%%%%%%%%%%%%%%%%%%%%%%%%%%%%%%%%%%%%%%%%%%%%%%%%%%%%%%%%%
\subsection{Motion Generator}
Our motion generator is a single Transformer encoder that receives input $\mathbf{U}$ and outputs transition frames $\mathbf{Y}_{tf}$ and the root destination vector $\mathbf{f}$.
The given frames are first projected to a latent space through a linear layer and then processed by the Transformer encoder.
The encoder follows the architecture of \citeN{vaswani2017attention}, consisting of multi-head attention and feed-forward layers.
It is composed of six Transformer encoder layers, with 8 heads and a hidden dimension of 512.
We adopt relative positional encoding \cite{huangmusic} without using absolute positional encoding, following prior work \cite{akhoundi2025silk, qin2022motion, hong2024long}. 
This allows the network to handle input sequences longer than the training samples.

\subsubsection{Adversarial training} \label{adv}
\begin{figure}
    \centering
    \includegraphics[width=0.4\textwidth]{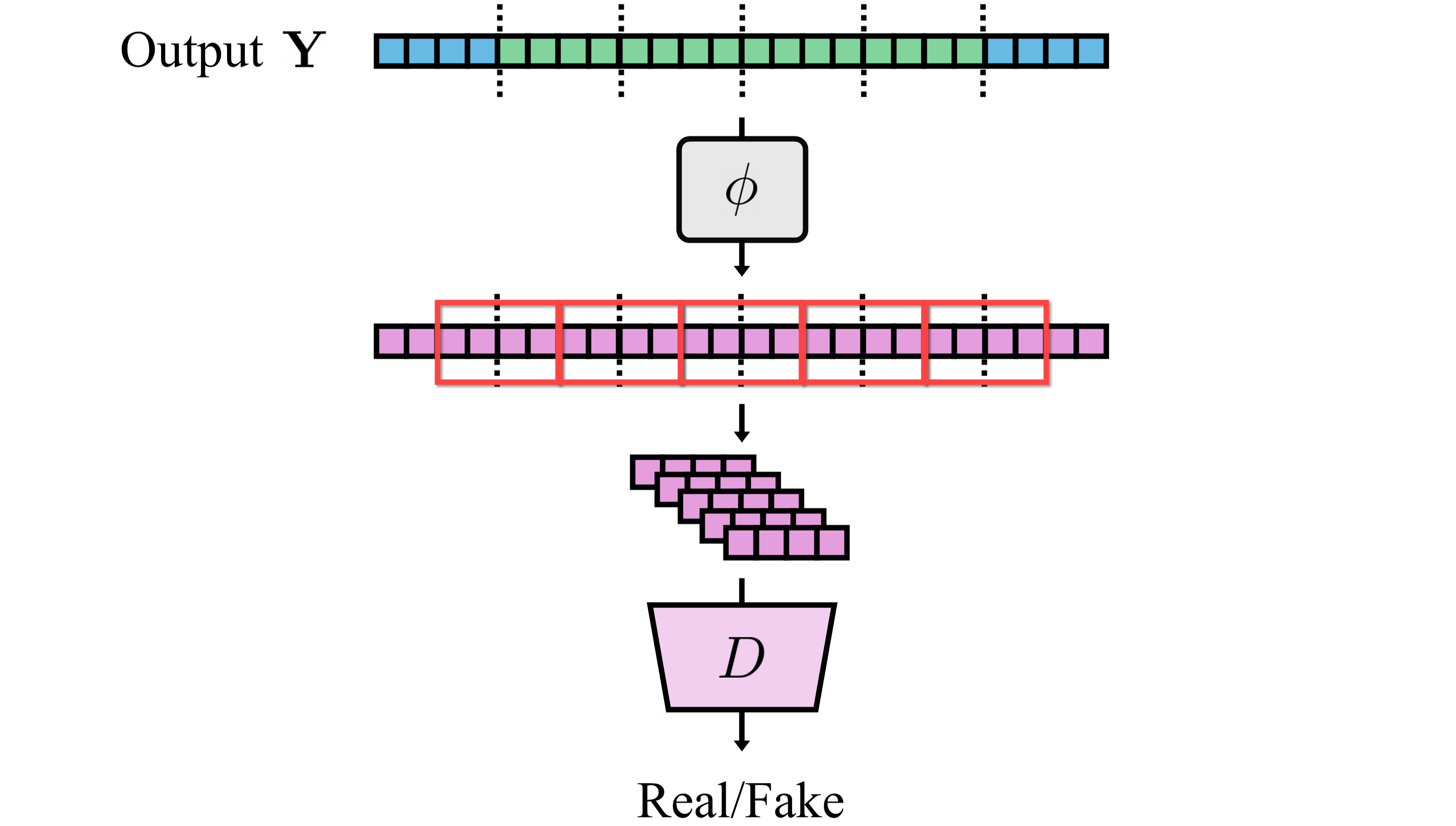}
    \caption{Data input to discriminator. 
    From the output sequence $\mathbf{Y}$, function $\phi$ extracts only the translational displacement.
    The dotted lines indicate the boundaries between blocks and the red boxes denote newly segmented motion blocks that serve as inputs to the discriminator $D$.}
    \label{fig:disc}
\end{figure}

Inspired by Harvey et al.~\shortcite{harvey2020robust}, we employ adversarial training for the motion generator.
The discriminator $D$ has two goals: (i) reducing popping artifacts at block boundaries and (ii) alleviating foot sliding across the generated sequence.
The input given to $D$ is illustrated in Figure~\ref{fig:disc}.
Given an output $\mathbf{Y}$, we compute the translational displacement of the head and $J_{vq}$ joints, through the function $\phi$.
Then we newly segment the sequence into blocks, including the boundary frames that originate from the data preprocessing stage.
We stack the segmented blocks together in a batch, and provide it to the discriminator.
The network is composed of three 1D convolutional layers to compress the temporal dimension, and two linear layers to produce the scores that discriminate between generated and ground-truth blocks.
The scores across all blocks in a batch are averaged to yield the discriminator output.

\subsubsection{Loss function}
% loss 설명
The total loss function for the motion generator is defined as follows:
\begin{align}
\mathcal{L}_{gen} 
&= \lambda_{trans}\mathcal{L}_{trans} \, + \, \lambda_{fk}\mathcal{L}_{fk} \notag \\
&\quad + \,  \lambda_{dest}\mathcal{L}_{dest}\, + \,\lambda_{adv}\mathcal{L}_{adv}, \notag
\end{align}
\noindent
where $\lambda_{trans}$, $\lambda_{fk}$, $\lambda_{dest}$, and $\lambda_{adv}$ are the weights that control the effect of each loss term. 
In the following paragraphs, we use $\widetilde{\phantom{x}}$ to denote ground truth values.

The transition loss compares the output transition poses with those of paired ground truth:
% Trans loss
\[
\mathcal{L}_{trans} = 
\frac{1}{T-2T_l} \sum_{t=T_l+1}^{T-{T_l}}\|\mathbf{x}_{t} - \tilde{\mathbf{x}}_{t}\|_1,
\]
\noindent
where we only compute for transition frames in the range $[T_l+1, T-T_l]$.

The forward \hm{kinematics (FK)} loss compute the joint position difference in the world frame:
% FK loss
\[
\mathcal{L}_{fk} = 
\frac{1}{T-2T_l} \sum_{t=T_l+1}^{T-T_l}
\|FK(\mathbf{T}_t, \mathbf{j}_t^{qrel}) - FK(\tilde{\mathbf{T}}_{t}, \tilde{\mathbf{j}}_{t}^{qrel})\|_1,
\]
\noindent
where $FK(\cdot)$ represents the process of converting root relative joint orientations $\mathbf{j}_t^{qrel}$ to parent relative orientations, and computing forward kinematics with the reconstructed root transformation $\mathbf{T}_t$.

The destination loss supervises the network to predict the root transformation at $T_{dest}$ by comparing $\mathbf{f}$ with the ground truth value:
% destination loss
\[
\mathcal{L}_{dest} = 
\|\mathbf{f} - \tilde{\mathbf{f}}\|_1.
\]

As explained in Section~\ref{adv}, we employ the loss terms from Least Square Generative Adversarial \hm{Network (LSGAN)} \cite{mao2017least}.
For the generator, we use:
% Adv loss - LSGAN 
\[
\mathcal{L}_{adv} = 
\frac{1}{2} \mathbb{E}_{\mathbf{U}\sim p_{\mathrm{U}}}
[
(D(\phi(G(\mathbf{U})))-1)^2
],
\]
\noindent
where $G$ denotes the Transformer that generates motion.
Accordingly, the loss for the discriminator is as follows:
\begin{align}
\mathcal{L}_{disc} = 
&\frac{1}{2} \mathbb{E}_{\tilde{\mathbf{Y}}\sim p_{\mathrm{Y}}}
[
(D(\phi(\tilde{\mathbf{Y}}))-1)^2
]
+
\frac{1}{2} \mathbb{E}_{\mathbf{U}\sim p_{\mathrm{U}}}
[
D(\phi(G(\mathbf{U})))^2
] \notag \\
&\quad + \,
\lambda_{R1}\mathbb{E}_{\tilde{\mathbf{Y}}\sim p_{\mathrm{Y}}}
\!\left[\left\|\nabla_{\mathbf{Y}_{d}} D(\mathbf{Y}_{d})|_{\mathbf{Y}_{d}=\phi(\tilde{\mathbf{Y}})}
\right\|_2^2\right].
\notag
\end{align}
The last term is the R1 regularization to stabilize GAN training \cite{mescheder2018training}, where gradient is computed with respect to $\mathbf{Y}_{d}$, a sequence of translational displacement converted from ground truth $\tilde{\mathbf{Y}}$.
For all the motion generator losses, we set weights $\lambda_{trans}$, $\lambda_{fk}$, $\lambda_{dest}$, and $\lambda_{adv}$, as 1.0, 1.0, 1.0, and 2.0 respectively. 
For the discriminator, we use $\lambda_{R1}=10.0$.

\section{Experiments}
In this section, we present motion stitching results, an ablation study, a scalability evaluation, and comparisons with prior methods.
For all experiments, we use our framework trained on LAFAN1 dataset~\cite{harvey2020robust}.
We excluded sequences with obstacles and mirrored the remaining sequences for data augmentation.
\hm{Following \citeN{harvey2020robust}}, Subject 5 is used for validation and the others for training.
In visualized results, blue characters denote input motions and green characters denote generated transitions,
and unless otherwise stated, raw output poses are visualized every 8 frames.
Further implementation details are provided in the Section C of the supplementary material.

\subsection{Motion Stitching on Diverse Motions}
\begin{figure}
    \centering
    \includegraphics[width=0.45\textwidth]{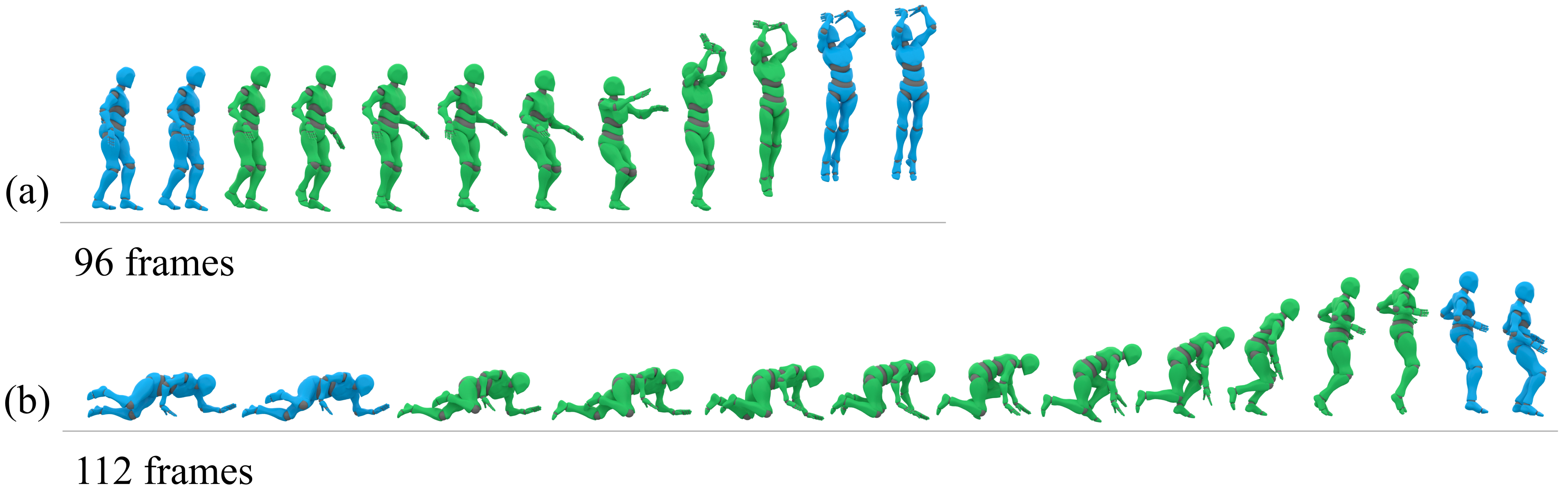}
    \caption{\hm{Motion stitching results on distinct motions. The total number of frames of output $\mathbf{Y}$ is given below each result.}}
    \label{fig:diverse_interm}
\end{figure}

We first demonstrate that our method effectively adapts the transition length according to the given motion pair.
As presented in Figure~\ref{fig:diverse_interm}, the character progresses through meaningful intermediate motions, such as preparing for a jump or stepping before lifting the body.
These results show that our method reaches the second motion through natural movements when given distinct motions as input.

\begin{figure}
    \centering
    \includegraphics[width=0.45\textwidth]{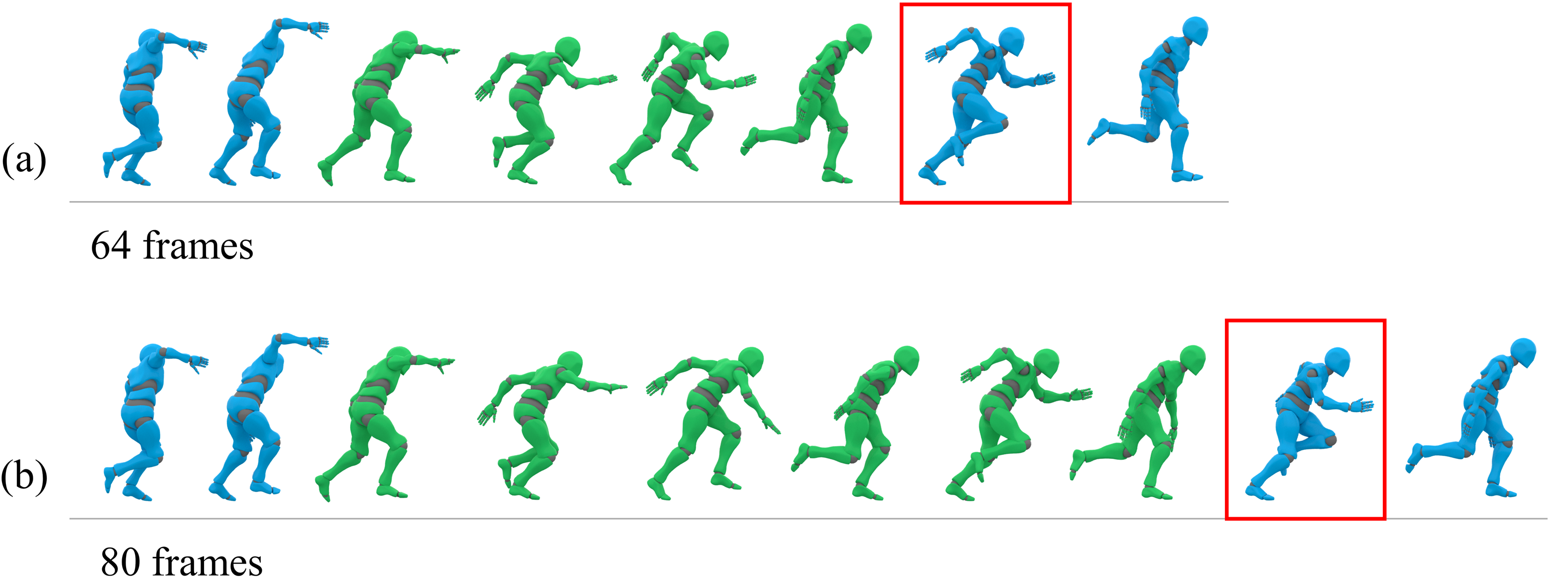}
    \caption{Stitching motions with difference foot phases. The first input motion is identical in both test cases. The red boxes mark the first footstep of the second input motion in each case.}
    \label{fig:diverse_phase}
\end{figure}

Our method also adapts the transition length based on foot phase.
As illustrated in Figure~\ref{fig:diverse_phase}, we compared two cases where the first input motion is identical but the second differs in foot phase.
When the second input begins with the left foot in \hm{contact (Figure~\ref{fig:diverse_phase} (a))}, the total length consists of 64 frames, whereas starting with the right \hm{foot (Figure~\ref{fig:diverse_phase} (b))} yields a longer motion with 80 frames by inserting an additional step.

\begin{figure}[b]
    \centering
    \includegraphics[width=0.47\textwidth]{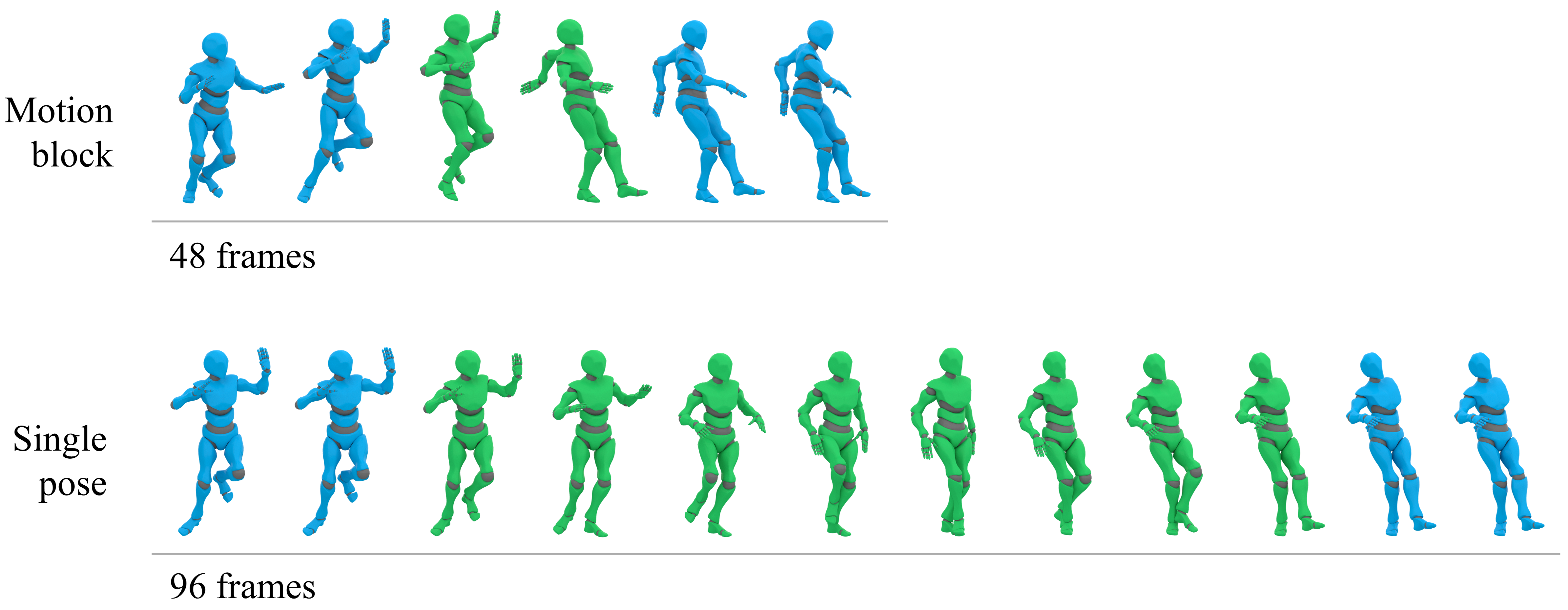}
    \caption{Comparison between the results from motion block and single pose inputs.}
    \label{fig:single_pose}
\end{figure}

When the two input clips contain only a single pose, our method interprets them as stationary motions with zero root velocity.
As shown in Figure~\ref{fig:single_pose}, we compared results obtained using full motion block inputs and repeated single-pose inputs.
For the latter, the boundary poses from the same motion pair are repeated to construct the input.
While motion block inputs produced quick transitions with varying root velocities, the single-pose case caused the character to gradually adjust its joint configurations toward the second pose, resulting in a longer transition.
Animated results are provided in the supplementary video (00:03\textasciitilde01:00).
%%%%%%%%%%%%%%%%%%%%%%%%%%%%%%%%%%%%%%%

\begin{figure*}
    \includegraphics[width=0.85\textwidth]{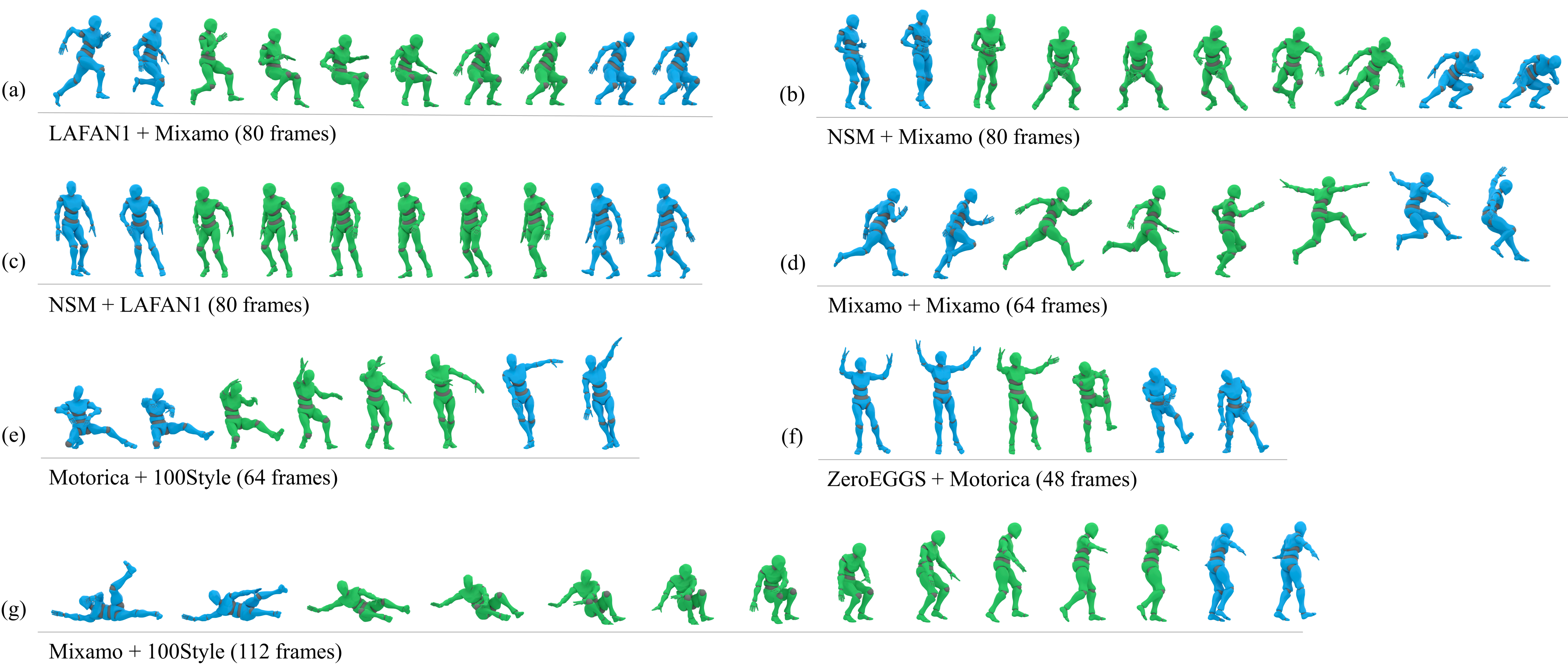}
    \caption{Motion stitching results on unseen motions. The source dataset of each input motion is indicated below each result, along with the total motion length.}
    \label{fig:unseen}
\end{figure*}

\begin{figure}
    \centering
    \includegraphics[width=0.38\textwidth]{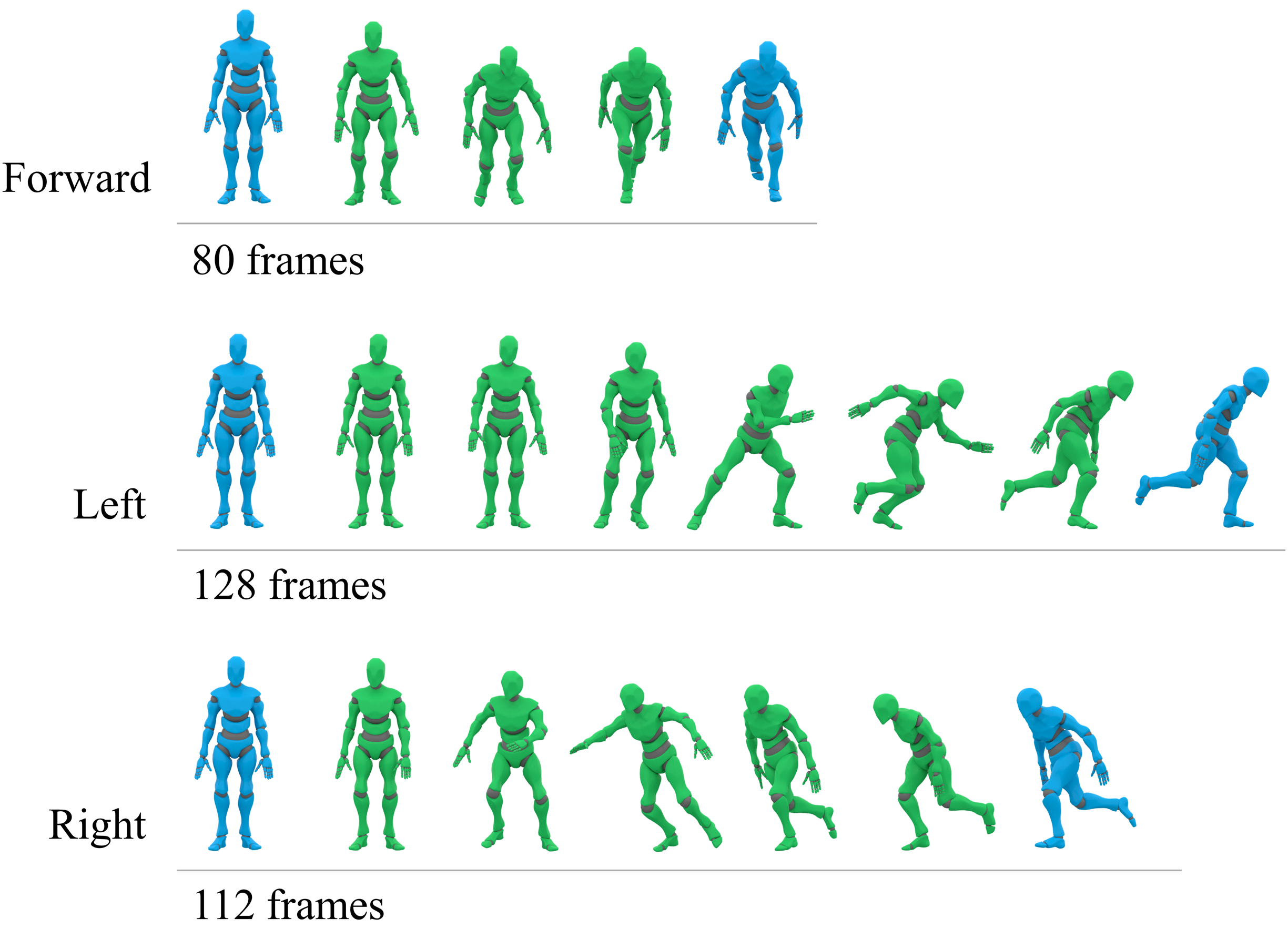}
    \caption{Controlling transition direction. Poses are visualized every 16 frames.}
    \label{fig:control}
\end{figure}

\subsection{Motion Stitching on Unseen Motions}\label{sec:unseen}
To assess how our method can be generalized to unseen motion sequences, we connected motions drawn from different datasets.
As shown in Figure~\ref{fig:unseen}, we tested combinations of sequences from LAFAN1, NSM~\cite{starke2019neural}, Mixamo~\cite{Mixamo}, ZeroEGGS~\cite{ghorbani2023zeroeggs}, Motorica~\cite{alexanderson2023listen,perez2021transflower}, and 100Style~\cite{mason2022local} without additional training.
The character successfully performed locomotive transitions, including slowing down to match a stationary \hm{pose (Figure~\ref{fig:unseen} (a))}, adapting foot placement between walking \hm{motions (Figure~\ref{fig:unseen} (b,c))}, and transitioning from sprinting to \hm{jumping (Figure~\ref{fig:unseen} (d))}.
Our method generated natural transitions between dance motion and stylized \hm{locomotion (Figure~\ref{fig:unseen} (e))}, or monologue \hm{gestures (Figure~\ref{fig:unseen} (f))}.
It further handled motions with rich ground contact, a characteristic commonly observed in sports \hm{sequences (Figure~\ref{fig:unseen} (g))}.
Animated results are presented in the accompanying video (01:03\textasciitilde01:49).

Overall, our method generated plausible transitions between two unseen motions.
We attribute this robustness to the use of predefined clusters and graph-based length estimation.
Each unseen motion is mapped to a known cluster, allowing the method to infer a plausible transition duration through pathfinding.
In addition, the guide input consists of cluster clips observed during training.
Consequently, the input to the motion generator remains consistent with the training distribution, enabling stable transition generation.
%%%%%%%%%%%%%%%%%%%%%%%%%%%%%%%%%%%%%%%

\subsection{Transition Direction Control}\label{control}
We demonstrate the flexibility of our approach through an additional experiment on direction control.
Because the motion generator is conditioned on the guide $\mathbf{S}_g$, the generation process can be steered by modifying $\mathbf{S}_g$ without retraining the network.
To enable directional control, we added a direction control term to both the cost \hm{function (Equation~\ref{eqn:cost})} and the heuristic \hm{function (Equation ~\ref{eqn:heu})} used during cluster pathfinding.
The control input was defined as a unit vector on the ground plane, representing the desired forward direction of the motion.
For each node visit, the direction control term was computed as the cosine distance between the given control input and root translational displacement in the world frame, accumulated from cluster clips.
The accumulated root displacement was also projected on the ground to compare the direction.
As a result, the cluster pathfinding selected nodes that guided the character to move in the specified direction.

Figure~\ref{fig:control} presents the results of this experiment.
All three examples use identical input motions, with control inputs of $[0, 0, 1]$ (forward), $[1, 0, 0]$ (left), and $[-1, 0, 0]$ (right).
As shown in the figure, the character successfully follows the specified control directions while matching the phase to the second motion.
In particular, the second and third results exhibit sharp directional turns because the modified cluster pathfinding selected direction changing movements.
These results demonstrate that our framework provides flexibility in transition generation by allowing the guide to be adapted using information from cluster clips.

%%%%%%%%%%%%%%%%%%%%%%%%%%%%%%%%%%%%%%%
%%%%%%%%%%%%%%%%%%%%%%%%%%%%%%%%%%%%%%%%%%%%%%%%%%%%%%%%%%%%%%%%%%%%%%%%%%%%%%%%%%%%%%%%%%%%%%%

\subsection{Ablation Study}
We conducted an ablation study on the guide input $\mathbf{S}_g$ and the destination loss $\mathcal{L}_{dest}$.
After training models under different conditions, we evaluated them on the motion generation test set using L2P and L2Q metric~\cite{harvey2020robust}, which measure the average L2 distances of the world positions and quaternions from the ground truth data, respectively.
We also provide comparisons of the raw generated motions to illustrate the visual impact of each component.

\subsubsection{Guide input}\label{sec:guide_input}
We evaluated the impact of providing $\mathbf{S}_g$ as an input to the motion generator.
For comparison, we trained an additional motion generator without $\mathbf{S}_g$, using only the target $\mathbf{X}_{tgt}$ as input.
This configuration resembles a motion in-betweening formulation~\cite{akhoundi2025silk}, where the model predicts transition frames given a set of dense context frames, zeroed out transition frames, and the last keyframe.
As shown in the first and last rows of Table~\ref{tab:ablation}, excluding the guide information caused the model to produce results that deviated from the ground truth data. 
This difference is also visible in the generated motions, as shown in the accompanying video (01:50\textasciitilde02:28).
Without direct initialization from $\mathbf{S}_g$, the character tended to slide when transitioning to the next input motion, instead of performing natural intermediate poses.
For example, the character's hip remains elevated for an extended period and the character rises abruptly without any movement to support its body from the ground.
These observations suggest that $\mathbf{S}_g$ helps disambiguate the generation process, effectively guiding the network to produce detailed transitions.
Additional results on the effect of $\mathbf{S}_g$ are provided in Section~D of the supplementary material.

\begin{table}[b]
\caption{Ablation study on guide and destination loss. The best result is highlighted in bold.}
\label{tab:ablation}
\centering
\begin{tabular}{lll}
\hline
Ablation & L2P $\downarrow$  & L2Q $\downarrow$            \\ \hline
No $\mathbf{S}_g$         & 0.567                               & 0.972         \\
% No $\mathbf{X}_{tgt}$         & 0.476                               & 1.110         \\
No $\mathcal{L}_{dest}$          & 0.442                               & 0.918          \\ \hline
Ours        & \textbf{0.436}                               & \textbf{0.909}         \\ \hline
\end{tabular}
\end{table}

{
\begin{table*}
\caption{Evaluation of graph scalability. $\mathcal{L}_{recon}$ denotes the validation reconstruction loss after CVQ-VAE training, and codebook size indicates the number of codes used for each dataset. Pathfinding time and graph memory are reported from the graph constructed with each codebook.}
\label{tab:scale_graph}
\centering
\begin{tabularx}{\linewidth}{l p{3.9cm} C{1cm} Y Y Y}
\hline
\\[-0.8em]
Dataset
& {\begin{tabular}[c]{@{}c@{}}Dataset size\\ (hours)\end{tabular}}
& {\begin{tabular}[c]{@{}c@{}}{$\mathcal{L}_{recon}$}\end{tabular}}
& {\begin{tabular}[c]{@{}c@{}}Codebook size \\ (dim.)\end{tabular}}
& {\begin{tabular}[c]{@{}c@{}}Pathfinding time \\ (sec)\end{tabular}}
& {\begin{tabular}[c]{@{}c@{}}Graph memory \\ (MB)\end{tabular}}
\\[-1.1em]
\\
% \midrule
\midrule
LAFAN1                       & 6.970 = 6.970                       & 1.933      & 2048  & 0.037  & 25.213 \\
LAFAN1 + InterAct              & 6.970 + 18.637 = 25.607              & 2.297      & 4096  & 0.074  & 82.273\\
LAFAN1 + InterAct + 100Style     & 6.970 + 18.637 + 44.257 = 69.864     & 1.733      & 8192  & 0.207  & 290.952 \\ \hline 
\end{tabularx}
\end{table*}
}

\subsubsection{Destination loss}
We conducted an ablation study on the destination loss $\mathcal{L}_{dest}$ to examine its effect on generating smoothly connected motion.
A motion generator was trained without $\mathcal{L}_{dest}$, producing only $\mathbf{Y}_{tf}$ without the root destination vector $\mathbf{f}$.
In this case, we aligned $\mathbf{X}_B$ to the last predicted root of $\mathbf{Y}_{tf}$ to obtain the final output $\mathbf{Y}$.
As shown in the second and the last rows of Table~\ref{tab:ablation}, our full model achieved results closer to the ground truth compared to the model that was trained without the loss.
The qualitative comparison is provided in the supplementary video (02:29\textasciitilde03:04).
The results generated without $\mathcal{L}_{dest}$ exhibited popping artifacts at the second boundary where the generated motion meets the second input motion, caused by the discontinuity in the root transformation.
In contrast, our method produced smooth transitions without such artifacts.
These findings emphasize the importance of predicting the root of the next frame to ensure seamless connections between the generated poses and the second input motion.

%%%%%%%%%%%%%%%%%%%%%%%%%%%%%%%%%%%%%%%
%%%%%%%%%%%%%%%%%%%%%%%%%%%%%%%%%%%%%%%%%%%%%%%%%%%%%%%%%%%%%%%%%%%%%%%%%%%%%%%%%%%%%%%%%%%%%%%

\subsection{Scalability}\label{sec:scale}
\subsubsection{Scalability of cluster transition graph}\label{sec:scale_graph}
As shown in Table~\ref{tab:scale_graph}, we examined the scalability of our cluster transition graph by increasing the size and diversity of the training dataset.
Starting from LAFAN1, we progressively added InterAct~\cite{huang2024interact} for single character and the 100Style dataset, and trained CVQ-VAE on the combined data with mirrored augmentation.
For each dataset configuration, we avoided using an excessively large codebook because it would reduce the average number of samples per cluster, potentially leading to an unbalanced training dataset for the motion generator.
Instead, we selected the codebook size at which further increasing it yielded only marginal improvements in reconstruction loss.
We then measured the pathfinding time for 1,000 randomly sampled start–goal node pairs.
We also report the memory usage for the graph, which consists of adjacency matrix, cost matrix for all possible node pairs, and cluster clips.
As presented in the table, the required codebook size for similar reconstruction performance increased only moderately as the dataset grows, while the other elements scaled with the codebook size. 
This indicates that CVQ-VAE effectively summarizes motion data from multiple sources into a finite set of codes, keeping the graph size manageable for practical use.

\subsubsection{Utilizing large dataset}
Our method can be extended to larger datasets than LAFAN1 that are not specifically designed for transition generation.
As shown in Table~\ref{tab:scale_method}, we trained a full pipeline $\mathrm{NMS}_{large}$, on a combined dataset of InterAct and 100Style without using LAFAN1.
Despite the substantial increase in dataset size, the codebook size modestly increased.
We attribute this to the characteristics of the datasets, where variation in root movement is relatively limited compared to LAFAN1, resulting in a relatively compact codebook.

\begin{figure}[b]
    \centering
    \includegraphics[width=0.45\textwidth]{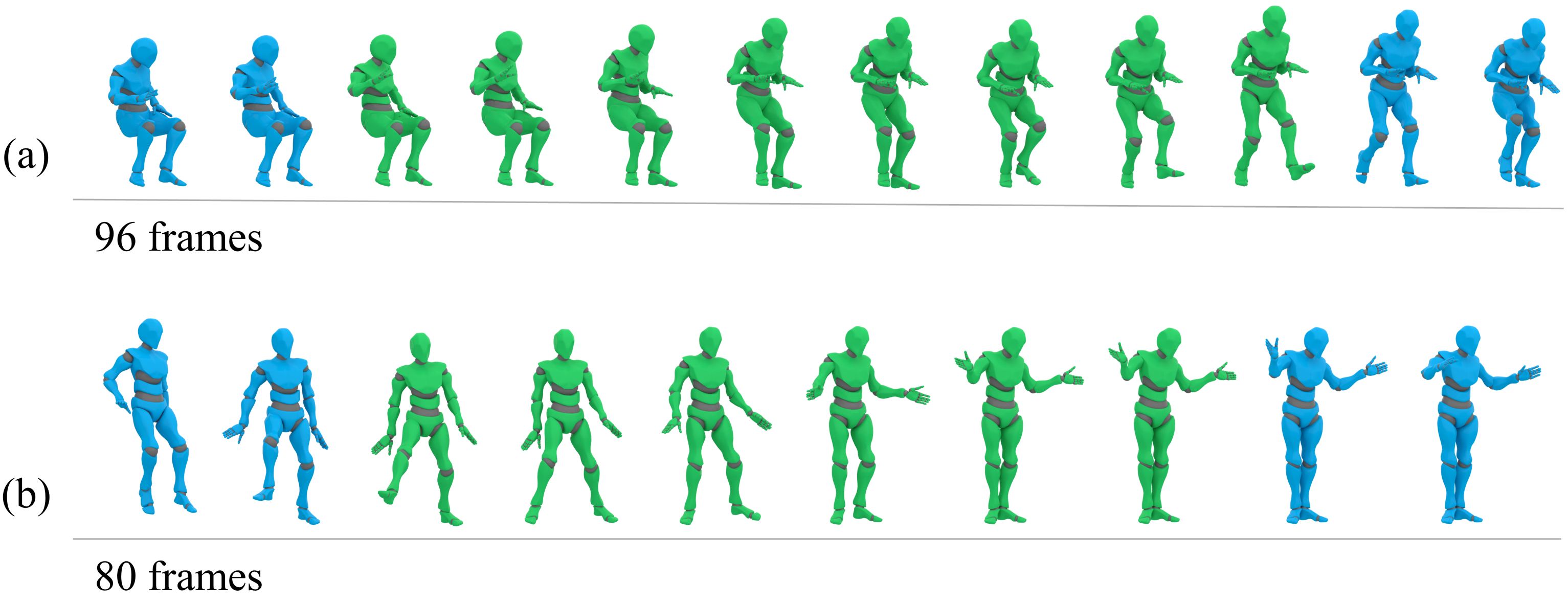}
    \caption{Qualitative results of $\mathrm{NMS}_{large}$.}
    \label{fig:scale_dataset}
\end{figure}

As demonstrated in Figure~\ref{fig:scale_dataset}, $\mathrm{NMS}_{large}$ successfully connects different input motions.
In each test case, the input pair is drawn from InterAct and 100Style, forming an unseen motion combination.
The model generates realistic transitions from sitting to dinosaur walking, and from kicking to shrugging.
These results indicate that our method can scale to larger datasets and learn effectively from combined motion sources.

{
\begin{table}[b]
\caption{Dataset used to train $\mathrm{NMS}_{large}$ compared with the original NMS. Dataset size in hours, CVQ-VAE reconstruction loss $\mathcal{L}_{recon}$, and the codebook size are also reported.}
\label{tab:scale_method}
\centering
\begin{tabularx}{\linewidth}{l Y Y Y Y}
\hline
\\[-1em]
Method
& {\begin{tabular}[c]{@{}c@{}}Dataset\end{tabular}}
& {\begin{tabular}[c]{@{}c@{}}Dataset\\size\end{tabular}}
& {\begin{tabular}[c]{@{}c@{}}{$\mathcal{L}_{recon}$}\end{tabular}}
& {\begin{tabular}[c]{@{}c@{}}Codebook\\size\end{tabular}}
\\[-1.3em]
\\ 
% \midrule
\midrule\\[-1.2em]
$\mathrm{NMS}$             & LAFAN1              & 6.970  & 1.934    & 2048\\[0.5em]
$\mathrm{NMS}_{large}$     & {\begin{tabular}[c]{@{}c@{}}InterAct\\+ 100Style \end{tabular}} & 62.894 & 1.379      & 4096\\[0.4em]
\hline 
\end{tabularx}
\end{table}
}

%%%%%%%%%%%%%%%%%%%%%%%%%%%%%%%%%%%%%%%%%%%%%%%%%%%%%%%%%%%%%%%%%%%%%%%%%%%%%%%%%%%%%%%%%%%%%%%

\subsection{Comparison}\label{sec:compare}
We compare our method to existing motion stitching methods, IB~\cite{bollo2018inertialization} and RMR~\cite{kim2023recurrent}, and motion in-betweening approaches, RMIB~\cite{harvey2020robust}, TS~\cite{qin2022motion}, NeMF~\cite{he2022nemf}, CondMDI~\cite{cohan2024flexible} \hm{and SILK~\cite{akhoundi2025silk}}.
The baselines are divided into two groups: fixed-length and variable-length methods.
By comparing with fixed-length methods, we show that generating varying transition lengths is crucial to obtain high-quality motion.
We further compare against variable-length methods that use direct transition time estimation and show the effectiveness of using a cluster transition graph for transition length estimation.
For completeness, we provide additional comparisons with another diffusion-based architecture in Section~E of the supplementary material. % for CAMDM

% dataset
\subsubsection{Test datasets}
We use three test datasets to compare our method with the two baseline groups.
The first dataset, $\mathrm{LAFAN1}_{within}$, consists of input pairs extracted within each LAFAN1 test sequence using the same sliding window size and stride described in Section~\ref{sec:mg_dataset}.
We defined the first and last motion blocks of 16 frames in a window as the two inputs, resulting in $66,780$ input pairs.
The second dataset, $\mathrm{LAFAN1}_{random}$, is constructed by sampling motion blocks from different LAFAN1 action categories.
We first group motion sequences according to their labels, then randomly select one input block from one category and the other from a different category, yielding 10,000 pairs. 
This dataset evaluates how well each method connects distinct motions.
The final dataset, Mixamo + 100Style, contains pairs where the first input motion is sampled from Mixamo and the second from 100Style. This dataset evaluates generalization to unseen motion sequences.

\begin{figure*}
    \centering
    \includegraphics[width=0.85\textwidth]{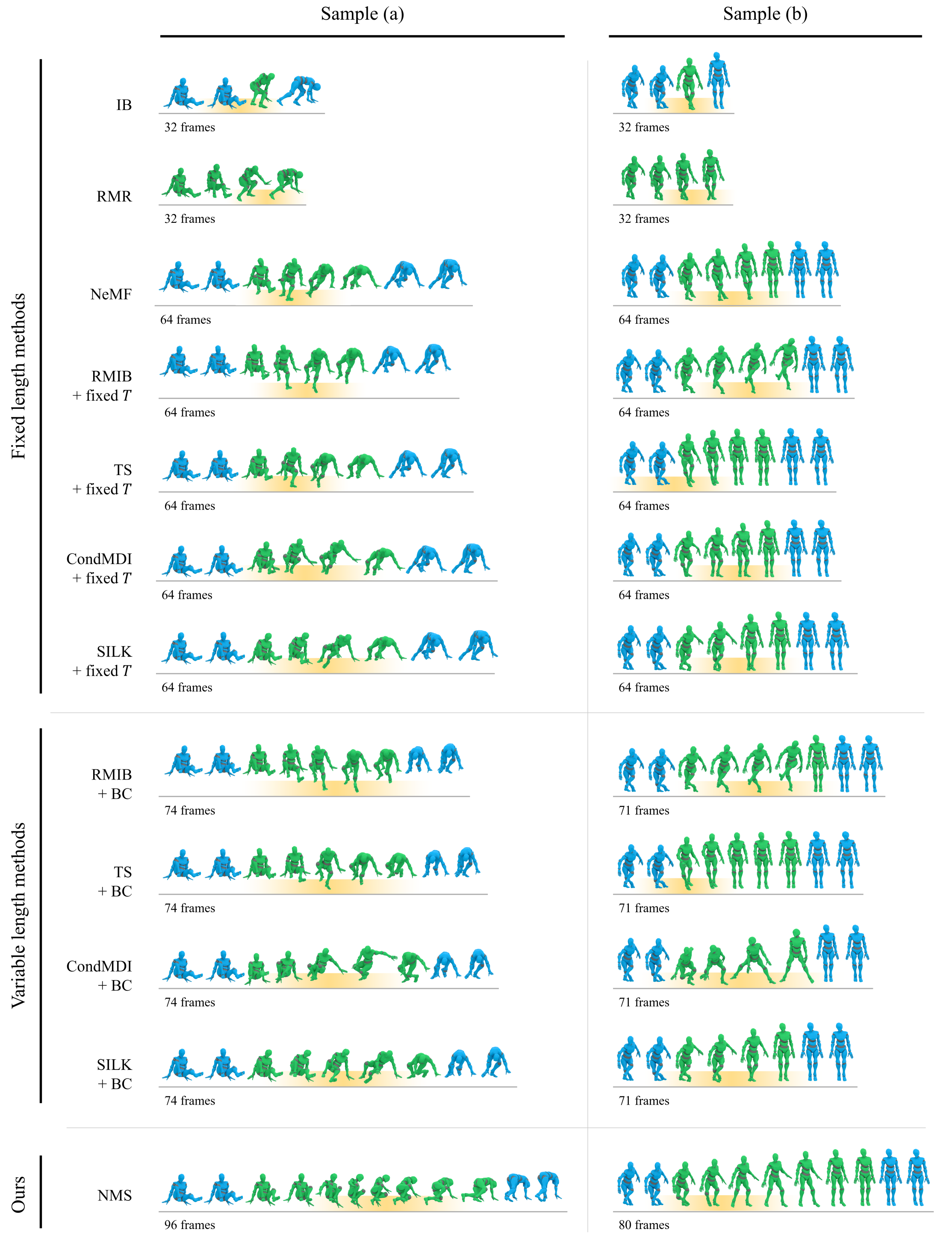}
    \caption{Qualitative comparison with other methods. In this figure, green characters indicate the poses edited or generated by each method. Notable poses in each result are highlighted in yellow.}
    \label{fig:compare_qual}
\end{figure*}

\begin{table*}
\centering
\small
\caption{Comparison of FID, detail distance, jitter, FS, and GP across different methods. The first row reports results on motion capture sequences. The second to seventh rows represent fixed-length methods, and the remaining rows represent variable-length methods. Best scores are shown in bold, and second-best scores are underlined.}
\label{tab:metric}
\begin{tabularx}{\textwidth}{
l
C{0.65cm} C{0.65cm} C{0.8cm} C{0.65cm} C{0.65cm}
C{0.65cm} C{0.65cm} C{0.8cm} C{0.65cm} C{0.65cm}
C{0.65cm} C{0.65cm} C{0.8cm} C{0.65cm} C{0.65cm}
}
\toprule
& \multicolumn{5}{c}{$\mathrm{LAFAN1}_{within}$}
& \multicolumn{5}{c}{$\mathrm{LAFAN1}_{random}$}
& \multicolumn{5}{c}{Mixamo + 100Style} \\
\cmidrule(lr){2-6}
\cmidrule(lr){7-11}
\cmidrule(lr){12-16}

Method
& \begin{tabular}[c]{@{}c@{}}FID$\downarrow$\end{tabular} 
& \begin{tabular}[c]{@{}c@{}}Detail\\dist.$\downarrow$\end{tabular} 
& \begin{tabular}[c]{@{}c@{}}Jitter$\downarrow$\end{tabular}
& \begin{tabular}[c]{@{}c@{}}FS$\downarrow$\end{tabular}
& \begin{tabular}[c]{@{}c@{}}GP$\downarrow$\end{tabular}
& \begin{tabular}[c]{@{}c@{}}FID$\downarrow$\end{tabular} 
& \begin{tabular}[c]{@{}c@{}}Detail\\dist.$\downarrow$\end{tabular} 
& \begin{tabular}[c]{@{}c@{}}Jitter$\downarrow$\end{tabular}
& \begin{tabular}[c]{@{}c@{}}FS$\downarrow$\end{tabular}
& \begin{tabular}[c]{@{}c@{}}GP$\downarrow$\end{tabular}
& \begin{tabular}[c]{@{}c@{}}FID$\downarrow$\end{tabular} 
& \begin{tabular}[c]{@{}c@{}}Detail\\dist.$\downarrow$\end{tabular} 
& \begin{tabular}[c]{@{}c@{}}Jitter$\downarrow$\end{tabular}
& \begin{tabular}[c]{@{}c@{}}FS$\downarrow$\end{tabular}
& \begin{tabular}[c]{@{}c@{}}GP$\downarrow$\end{tabular} \\
\midrule
Mocap & 0.01 & -- & 273.38 & 0.12 & 0.09 & -- & -- & -- & -- & -- & -- & -- & -- & -- & --\\
\midrule
IB & 1.17 & 11.85 & 920.18 & 1.16 & 0.11 & 9.27 & 40.69 & 1827.77 & 1.80 & 0.13 & 18.45 & 43.95 & 1563.18 & 1.78 & \textbf{0.04} \\
RMR & 1.55 & 14.00 & \underline{689.36} & 0.33 & \textbf{0.09} & 10.34 & 46.28 & 1875.18 & 0.68 & 0.20 & 20.92 & 74.25 & 2456.11 & 0.94 & 0.12 \\
NeMF & 0.38 & 12.85 & 1189.86 & 0.54 & 0.15 & 8.29 & 22.25 & 1488.02 & 0.99 & 0.22 & 15.88 & 33.36 & 1495.07 & 1.37 & 0.11 \\
RMIB + fixed $T$ & 0.36 & 15.42 & 1118.32 & 0.59 & 0.14 & 7.97 & \underline{15.51} & 1492.89 & 0.93 & 0.22 & 19.04 & 33.71 & 4174.35 & 4.69 & 0.11 \\
TS + fixed $T$ & \underline{0.30} & 13.33 & 1039.67 & \underline{0.27} & \textbf{0.09} & \underline{7.61} & \textbf{13.66} & 1252.96 & 0.75 & \underline{0.10} & \underline{14.02} & \underline{26.14} & 2676.78 & 0.90 & \underline{0.05} \\
CondMDI + fixed $T$ & 0.58 & \underline{7.29} & 3797.48 & 0.54 & \textbf{0.09} & 8.05 & 15.90 & 4510.02 & 1.14 & \textbf{0.09} & 14.58 & 31.56 & 5131.86 & 1.46 & 0.06 \\
SILK + fixed $T$ & 0.56 & 10.45 & 1154.88 & 0.85 & \underline{0.10} & 8.24 & 23.57 & 1380.73 & 1.28 & 0.11 & 14.30 & 28.52 & 1354.38 & 1.41 & \underline{0.05} \\
\midrule
RMIB + BC & 0.36 & 13.61 & 1302.39 & 0.77 & 0.15 & 7.95 & 24.78 & 1404.16 & 0.74 & 0.27 & 21.81 & 45.96 & 3179.67 & 3.72 & 0.22 \\
TS + BC & 0.34 & 11.67 & 1235.63 & 0.28 & \textbf{0.09} & 7.66 & 20.02 & \underline{1016.50} & \underline{0.44} & 0.13 & 14.54 & 39.79 & 1774.01 & \underline{0.60} & 0.08 \\
CondMDI + BC & 4.03 & 32.22 & 10656.29 & 1.88 & 0.14 & 20.61 & 58.78 & 17456.14 & 3.53 & 0.22 & 30.33 & 53.08 & 19912.32 & 4.12 & 0.16 \\
SILK + BC & 0.44 & 8.63 & 1284.75 & 1.09 & \underline{0.10} & 7.91 & 18.21 & 1139.37 & 0.97 & 0.11 & \textbf{12.51} & \textbf{23.26} & \underline{1000.23} & 0.90 & 0.07 \\
\midrule
NMS (Ours) & \textbf{0.20} & \textbf{5.18} & \textbf{382.89} & \textbf{0.23} & \textbf{0.09} & \textbf{7.60} & 15.86 & \textbf{385.00} & \textbf{0.38} & \underline{0.10} & 14.26 & 26.54 & \textbf{589.82} & \textbf{0.52} & \underline{0.05} \\
\bottomrule
\end{tabularx}
\end{table*}

% baselines
\subsubsection{Baselines}\label{sec:baselines}
The fixed-length methods include motion stitching methods IB and RMR, and motion in-betweening methods with fixed total length $T$.
Given the two input motions with 16 frames each, we applied each method and obtained connected motion sequences.
For IB, we aligned the two inputs and blended the poses for 10 frames at the boundary, producing a sequence of 32 frames.
For RMR, it receives aligned inputs and produces a connected motion with the same length, generating the entire sequence of 32 frames.
For NeMF, RMIB, TS, CondMDI \hm{and SILK}, we generated 32 in-between poses and inserted them between the two inputs, producing a motion with $T=64$ frames.
This fixed length was chosen by computing the mean output length of our method across the $\mathrm{LAFAN1}_{within}$, and selecting the closest supported length.
All neural network-based methods were trained with training sequences in LAFAN1 with 60 FPS, and tested on the three datasets without additional training.

For variable-length methods, we evaluated our method against blend length \hm{computation (BC)}~\cite{wang2008synthesis} combined with motion in-betweening models.
Following Wang et al.~\shortcite{wang2008synthesis}, the transition duration was computed based on joint angular displacement, where the last frame of the first input and the first frame of the second input were treated as the start and end of the transition, respectively.
Using the estimated transition time, we approximated the target keyframe and generated poses using the motion in-betweening methods.
Among the models categorized as fixed-length methods, we selected those capable of generating variable-length outputs, namely RMIB, TS, CondMDI, \hm{and SILK}.
These combined methods produced output motions ranging in total length from 33 to 212 frames.

In the case of motion in-betweening methods, we defined all 16 frames in the first input as the context frames, and the first frame of the second input as the target keyframe.
Because we assume that both inputs are aligned to the identity transformation in the world frame, we estimated the world transformation of the target keyframe through a separate process.
Given the number of transition frames, we interpolated the root translational and angular displacement at the last frame of the first input and the first frame of the second input. 
We then accumulated the interpolated displacements to obtain a root transformation for the target keyframe.

\subsubsection{Qualitative comparison}
Qualitative comparisons are shown in Figure~\ref{fig:compare_qual}. The first sample is from \(\mathrm{LAFAN1}_{random}\), and the second is from the Mixamo + 100Style dataset. 
Animated comparisons are provided in the supplementary video \hm{(03:21\textasciitilde04:47)}.
As illustrated in the last row, our method generated gradual intermediate poses that guided the character to the second input.
Among fixed-length methods, IB produced abrupt pose changes, and RMR struggled to preserve the characteristics of the original input motions.
The other motion in-betweening methods produced better results than IB and RMR by inserting new intermediate poses.
However, they tended to generate overly smoothed motions for challenging pairs such as getting up in Sample (a), and exhibited boundary discontinuities or implausible poses for unseen inputs, as shown in Sample (b).

As shown in the last four rows of Figure~\ref{fig:compare_qual}, combining BC with motion in-betweening methods often led to unnatural results.
When BC underestimated transition time, as in Sample (a), RMIB, TS, CondMDI, \hm{and SILK} generated abrupt pose changes, whereas our method allocated sufficient time for the character to get up with limbs supporting the body.
Even with similar predicted lengths, as in Sample (b), our method achieved higher motion fidelity compared to the BC-based methods, producing smooth transition for unseen input motions without discontinuities.

\subsubsection{Quantitative comparison}\label{sec:metric}
As no ground truth motion exists for the shortest transition, we evaluated performance using Frechet Inception \hm{Distance (FID)}, motion detail distance, jitter, \hm{foot sliding (FS), and ground penetration (GP)}.
We computed FID using an RNN-based autoencoder~\cite{guo2024momask} to measure how closely the generated local pose sequences resemble motion capture data.
We also reported motion detail distance, inspired by motion detail metric by~\citeN{starke2022deepphase}.
The original metric measures the vividness of generated motions by taking the average of joint rotations per second, where higher values indicate greater motion detail.
However, in our setting, this can overestimate quality for motions that move too rapidly within a limited duration.
Therefore, we compute the Wasserstein distance between the distribution of detail values from generated sequences and that of motion capture sequences.
This penalized both overly static motions and excessively abrupt movements, better reflecting overall motion quality.
We use jitter~\cite{yi2021transpose}, which quantifies motion smoothness by measuring the average joint jerk.
In our experiments, jitter mainly captures how smoothly the generated transition connects the inputs at the boundary.
\hm{To evaluate foot-contact quality, we computed FS~\cite{zhang2018mode} and scaled the values by $10^{3}$.}
\hm{Finally, we computed GP~\cite{lee2023same} and reported the values in centimeters.}
Note that, for reference, we included scores from test sequences in \hm{LAFAN1 (Mocap)} when reporting results on $\mathrm{LAFAN1}_{within}$.

The resulting metric scores are reported in Table~\ref{tab:metric}. 
Among fixed-length methods, NeMF, RMIB, TS, CondMDI, \hm{and SILK} outperformed IB and RMR in terms of FID and detail distance on $\mathrm{LAFAN1}_{random}$ and Mixamo + 100Style, indicating their ability to generate plausible poses.
However, these methods exhibited high jitter, which we attribute to discontinuities at the boundaries between input and generated motions.
In particular, TS with fixed $T$ achieved strong FID and detail distance scores but suffered from high jitter. 
This suggests that TS struggles to match boundary poses for unseen inputs, which is consistent with the qualitative results presented in Figure~\ref{fig:compare_qual}.
In contrast, our method achieved the best scores for $\mathrm{LAFAN1}_{within}$, and significantly lower jitter for $\mathrm{LAFAN1}_{random}$ and Mixamo + 100Style, while maintaining comparable FID and detail distance.

\begin{figure}[t]
    \centering
    \includegraphics[width=0.47\textwidth]{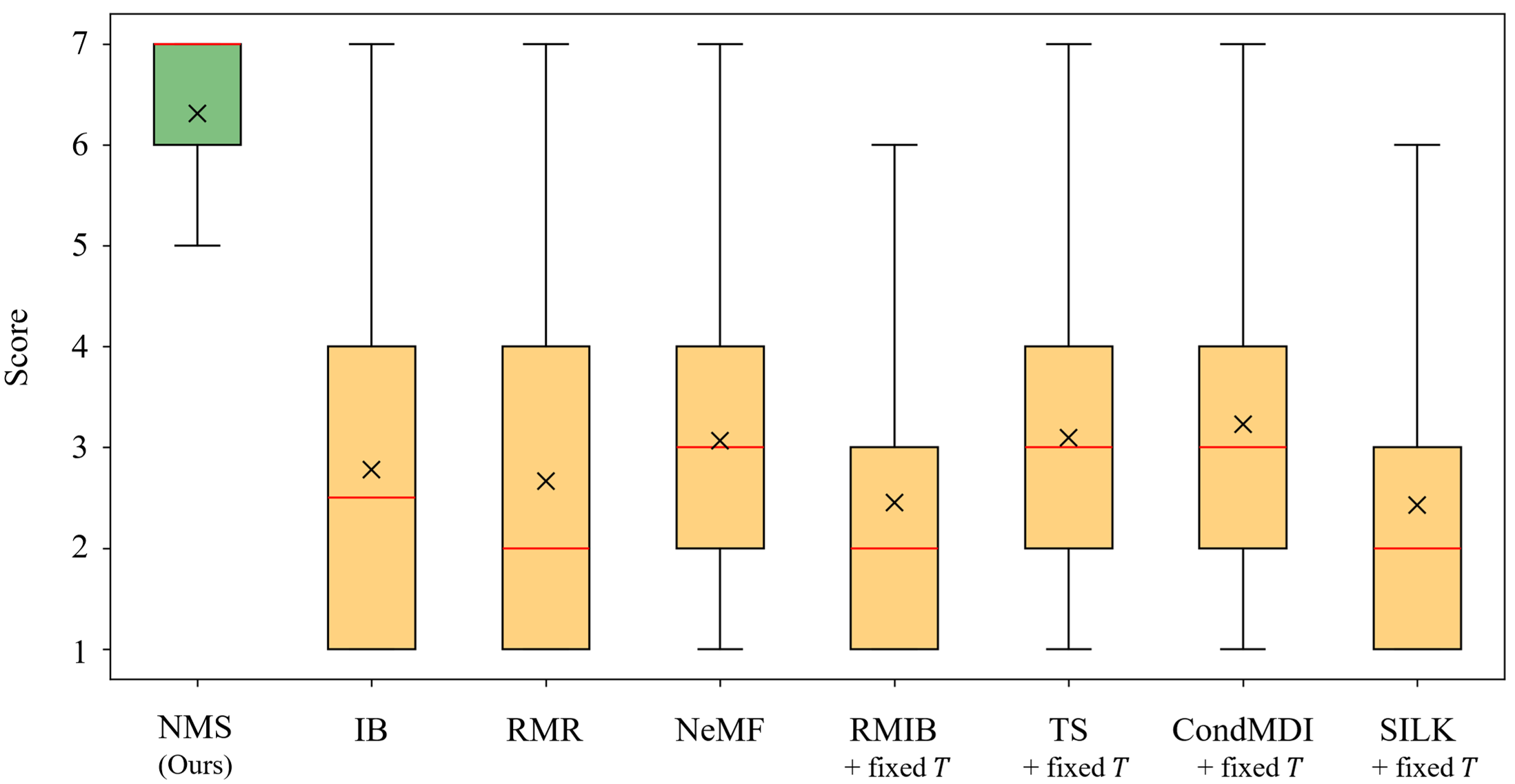}
    \caption{\hm{User study results shown as box plots.}}
    \label{fig:userstudy}
\end{figure}

As shown in the last \hm{five} rows in Table~\ref{tab:metric}, combining BC with motion in-betweening methods did not yield significant performance improvements.
This suggests that BC failed to estimate realistic transition lengths required for optimal performance.
\hm{Our method outperformed all BC-based approaches except SILK + BC on Mixamo + 100Style.
SILK tended to generate motions close to the training distribution, producing poses that resemble motion-capture sequences.
However, such poses do not necessarily yield perceptually natural transitions for unseen input motions, as reflected in the jitter, FS, GP, and qualitative results.
Our method achieved strong performance on these criteria.
We attribute its overall performance advantage to graph-based transition length estimation.}
By traversing the graph, our method implicitly captures the time required for specific movement patterns, whereas BC estimates the length solely from the joint angular displacement at the boundary.

\hm{
\subsubsection{User study}
We conducted a user study to further evaluate the visual quality of the results produced by our method relative to fixed-length methods with comparable quantitative performance.
The study involved 20 participants aged 23 to 36.
Participants rated their familiarity with character animation on a 7-point scale, where the average rating was 4.9, indicating that the participants had moderate familiarity with character animation.
Then they rated the naturalness of the generated motion sequences on a 7-point Likert scale across 10 test cases~\cite{zhang2025motion}.
The test set consists of five locomotion scenarios and five challenging cases, such as transitions from lying down to jumping.
The results shown in Figure~\ref{fig:userstudy} suggest that our method provides strong visual quality, receiving the highest score across methods.
This difference is likely due to the ability of our approach to produce gradual transitions that remain consistent with the input motions.
Among the baselines, NeMF, TS, and CondMDI attained higher mean scores than the others, as they generated more plausible poses on unseen inputs than those from IB, RMR, RMIB, and SILK.
}

\section{Discussion}\label{discussion}
\subsection{Effect of Training Dataset}
Although our method successfully generates gradual transition poses, it can still produce popping artifacts at the first boundary when the input motions deviate from the training distribution.
Empirically, we found that this is more likely to occur when the character begins in relatively rare poses without two-foot \hm{support (e.g. crawling, kicking, or lying down)}, or when the input motion originates from an unseen dataset.
While such artifacts can affect visual quality, they can be effectively mitigated through simple post-process blending.
A comparison of sequences obtained with and without postprocessing is provided in the supplementary video \hm{(04:47\textasciitilde04:55)}.

To achieve optimal performance, our method requires high-quality motion clusters constructed from training datasets that include diverse motion capture sequences.
\hm{If the data is limited in quantity or diversity, an input motion may be mapped to an unsuitable cluster.
This can produce incorrect transitions even between semantically plausible input motions, such as running to walking.}
For example, as shown in \hm{Figure~\ref{fig:limitation} (a)}, when our framework trained on LAFAN1 is given diving and walking motions as input, it struggles to generate realistic outputs because there \hm{are} no motion clusters related to swimming sequences.
We expect that including a diverse set of motions in the dataset will further expand the coverage of motion clusters and improve realism across a wider range of input combinations.

\subsection{Pathfinding-based Guide Generation}
{
\begin{table}
\centering
\caption{Inference time comparison in milliseconds.}
\label{tab:infer_time}
\setlength{\tabcolsep}{4pt}
\begin{tabularx}{0.9\columnwidth}{l *{3}{Y}}
\toprule
Method & \begin{tabular}[c]{@{}c@{}}Length\\estimation\end{tabular} & \begin{tabular}[c]{@{}c@{}}Motion\\generation\end{tabular} & Total \\
\midrule
IB                  & --  & 1.5     & 1.5 \\
RMR                 & --  & 2.6     & 2.6 \\
NeMF                & --  & 22007.5 & 22007.5 \\
RMIB + fixed $T$    & --  & 45.7    & 45.7 \\
TS + fixed $T$      & --  & 7.1     & 7.1 \\
CondMDI + fixed $T$ & --  & 15173.6 & 15173.6 \\
SILK + fixed $T$ & --  & 6.5 & 6.5 \\
\midrule
RMIB + BC           & 0.7 & 50.3    & 51.0 \\
TS + BC             & 0.7 & 6.5     & 7.2 \\
CondMDI + BC        & 0.7 & 14223.6 & 14224.3 \\
SILK + BC        & 0.7 & 6.5 & 7.2 \\
\midrule
NMS (Ours)          & 6.3 & 8.2     & 14.5 \\
\bottomrule
\end{tabularx}
\end{table}
}

%%%%%%%%%%%%%%%%%%%% A* algorithm
While cluster pathfinding serves as a core component of the method, it is deterministic and does not consider semantic conditions.
Unlike generative approaches that produce variations or support text and action conditioning, our method yields the same transitions when the cost settings are fixed.
However, using the pathfinding algorithm has advantages in that it requires no additional training and guarantees that a valid path to the goal node will be found.
It also allows for simple adjustments of cost terms and can produce diverse transitions from the same inputs, as demonstrated in Section~\ref{control}.

%%%%%%%%%%%%%%%%%%%% A* tuning issue
An important consideration when employing pathfinding algorithm is the need for manual tuning of the weight parameters.
Although the graph is constructed from valid transitions between clusters, the selected paths should resemble those seen during motion generator training so that the corresponding guide inputs remain consistent with the training samples.
We observed that the guides producing plausible results contained root movements that were smoother than those of other joints.
Therefore, as discussed in Section~\ref{traversal}, we prioritized root-related features to minimize artifacts.
\hm{Additional experiment on the pathfinding weight parameters is provided in Section F of the supplementary material.}

\hm{\subsection{Realtime Application}}
Because motion stitching is often used in real-time applications, the inference time of our method can be a limitation when A* is employed.
To evaluate the computational efficiency, we measured the inference time of our method and baseline methods on 1,000 test inputs sampled from $\mathrm{LAFAN1}_{within}$.
The average results are reported in Table~\ref{tab:infer_time}.
NeMF and CondMDI required approximately 22 and 15 seconds, respectively, due to latent-space optimization and iterative denoising.
Among the remaining baselines, IB and RMR were the fastest but generally produced low-quality motions, while RMIB-based methods were slower than TS-based \hm{or SILK-based} methods because they generate frames autoregressively.
When compared with TS-based \hm{and SILK-based} methods, the main bottleneck of our method was the length estimation via A*.
To improve efficiency, future work could explore replacing graph and search algorithm entirely with a neural network, producing guides that imitate the ones extracted from motion capture data.
Such a design could reduce the cost of length estimation while also automating the weight tuning.

\hm{
In addition, extending motion stitching to environment-aware scenarios can be an important direction for future work, particularly for real-time applications in which characters interact with surroundings. 
Our current framework focuses on transition generation in an empty environment. 
Incorporating terrain and object information similar to PFNN~\cite{holden2017phase} or NSM~\cite{starke2019neural}, could broaden its applicability to interactive settings. 
Such information could also be integrated into the pathfinding objective through interaction-aware cost terms, allowing the method to select motion states that respond appropriately to the surroundings while maintaining natural transitions.
}

\begin{table}[b]
\centering
\small
\caption{Comparison of input motion dissimilarity and motion quality across motion-length ranges. Length is measured by the total number of frames, and the best scores are shown in bold.}
\label{tab:length}
\begin{tabularx}{0.95\columnwidth}{
l
*{7}{C{0.6cm}}
}
\toprule
& \multicolumn{2}{c}{Input diff.}
& \multicolumn{5}{c}{Metrics} \\ 
\cmidrule(lr){2-3}
\cmidrule(lr){4-8}

Length
& \begin{tabular}[c]{@{}c@{}}Cosine\\dist.\end{tabular} 
& \begin{tabular}[c]{@{}c@{}}Pose\\diff.\end{tabular} 
& \begin{tabular}[c]{@{}c@{}}FID$\downarrow$\end{tabular} 
& \begin{tabular}[c]{@{}c@{}}Detail\\dist.$\downarrow$\end{tabular} 
& \begin{tabular}[c]{@{}c@{}}Jitter$\downarrow$\end{tabular}
& \begin{tabular}[c]{@{}c@{}}FS$\downarrow$\end{tabular}
& \begin{tabular}[c]{@{}c@{}}GP$\downarrow$\end{tabular} \\
\midrule
$[48, 80)$ & 0.45 & 0.41 & \textbf{1.63} & \textbf{7.34} & 422.46 & \textbf{0.25} & \textbf{0.08}\\
$[80, 128)$ & 0.75 & 0.69 & 5.85 & 12.05 & 402.97 & 0.26 & 0.09 \\
$[128, 176]$ & 0.80 & 1.83 & 12.09 & 17.89 & \textbf{374.95} & 0.75 & 0.14 \\
\bottomrule
\end{tabularx}
\end{table}

\hm{
\subsection{Variation of Transition Lengths}
Although our method adapts transition length according to input motion dissimilarity, motion quality may degrade in challenging cases.
To examine the relationship among transition length, input motion dissimilarity, and motion quality, we collected all generated motions from the three test datasets discussed in Section~\ref{sec:compare}, grouped them into three length-based bins, and randomly selected 500 samples from each bin.
For each bin, we measured the average cosine distance between the cluster codes of the input motions and the average boundary pose difference using Equation~\ref{eqn:bnd}.
We also computed all motion quality metrics introduced in Section~\ref{sec:metric} for all bins.
}

\hm{
As shown in Table~\ref{tab:length}, the output motion length generally increased with input dissimilarity.
The corresponding quality metrics also increased with output motion length, indicating a decline in motion quality.
This decline is likely associated with the difficulty of the input pairs, as the longest bin contains challenging transitions such as lying down to aiming and kneeling to running.
These transitions remain plausible as demonstrated by similar cases in Figure~\ref{fig:diverse_interm} (b) and Figure~\ref{fig:unseen} (g), but may occasionally lose fine motion details or exhibit foot sliding. 
To mitigate these artifacts, future work could incorporate an additional module that adapts the pathfinding weights to each input pair and encourages gradual pose changes in challenging cases.
}

\subsection{Graph Scalability}
%%%%%%%%%%%%%%%%%%%% Scaling up graph
As shown in Section~\ref{sec:scale}, the cluster transition graph can be scaled to datasets larger than LAFAN1 to some extent, but it still faces scalability challenges as graph memory usage grows with the training dataset size.
In the future, graph memory could be reduced by pruning rarely used edges or using a hierarchical graph structure. 
Nevertheless, the results presented in Section~\ref{sec:unseen} suggest that our method can generalize to unseen motion combinations when trained with sufficiently diverse datasets.
By providing the motion generator with appropriate transition durations and guides, the cluster transition graph helps produce plausible results even with a moderately sized motion dataset.

\subsection{Physical Plausibility}
\begin{figure}
    \centering
    \includegraphics[width=0.43\textwidth]{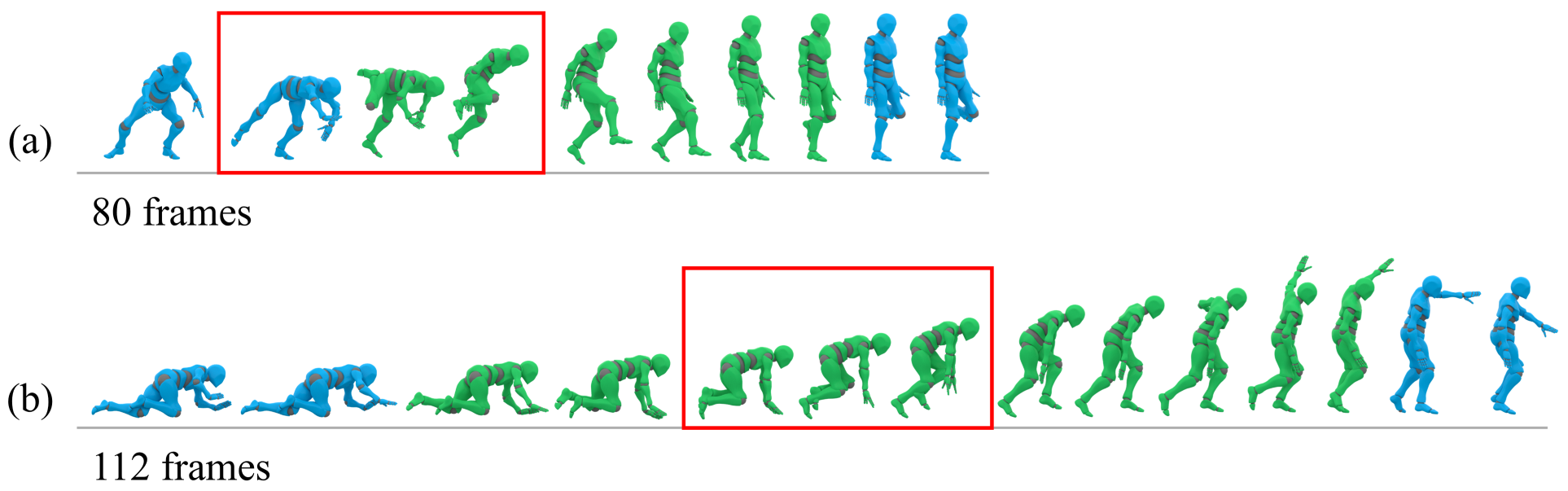}
    \caption{Failure cases of our method. (a) Input mapped to an unsuitable cluster. (b) Physically implausible motion. 
    The red boxes highlight unnatural poses.}
    \label{fig:limitation}
\end{figure}

Our method may produce results that do not fully comply with real-world physical principles because it does not incorporate any physics constraints.
When transitioning between motions that differ significantly, how the character supports the body with its limbs, or how the character's center of mass changes are critical for achieving high-fidelity motion.
Our approach captures these aspects to some extent by constraining the motion generation through graph traversal.
However, as shown in Figure~\ref{fig:limitation} (b), it can still produce smoothing artifacts such as the character lifting its hips without appropriate support from other body parts.
To further improve realism, physics-based terms could be integrated to model the relationship between contact forces and body movement, or physics simulation could be applied to refine the transitions.

\section{Conclusion}
In this work, we propose a neural network-based motion stitching framework that generates transitions of varying lengths using a cluster transition graph.
Our approach categorizes input motions, identifies intermediate motion states through the cluster pathfinding, and produces naturally connected transitions based on the discovered path and the given motions.
As demonstrated in our experiments, leveraging the cluster path as a motion guide reduces ambiguity during transition generation and effectively connects a wide range of motion types.
Because the guide is determined independently through cluster pathfinding, the motion generation process can be controlled by modifying the terms used in the pathfinding algorithm, without any additional training.
Comparisons with existing methods further show that generating varying-length transitions and inferring transition duration implicitly through graph structure improve motion stitching performance.
In the future, replacing the pathfinding module with a learnable path generator could eliminate manual weight tuning and enable diverse conditioning signals.
Other future directions include utilizing a diverse and well-balanced dataset to improve output quality, and incorporating physics-based constraints to enhance the realism of generated motions.

\begin{acks}
We thank Nicolas Nghiem for his invaluable help with rendering the results. This research was supported by Culture, Sports and Tourism R\&D Program through the Korea Creative Content Agency grant funded by the Ministry of Culture, Sports and Tourism in 2024 (RS-2024-00440434).
\end{acks}

\bibliographystyle{ACM-Reference-Format}
\bibliography{reference}

@String{Computing = "Computing" }

@String{Computer = "{IEEE} Computer" }

@String{Springer = "Springer-Verlag" }

@article{cho2021motion,
  title={Motion recommendation for online character control},
  author={Cho, Kyungmin and Kim, Chaelin and Park, Jungjin and Park, Joonkyu and Noh, Junyong},
  journal={ACM Transactions on Graphics (TOG)},
  volume={40},
  number={6},
  pages={1--16},
  year={2021},
  publisher={ACM New York, NY, USA}
}

@article{van2017neural,
  title={Neural discrete representation learning},
  author={Van Den Oord, Aaron and Vinyals, Oriol and others},
  journal={Advances in neural information processing systems},
  volume={30},
  year={2017}
}

@inproceedings{huangmusic,
  title={Music Transformer: Generating Music with Long-Term Structure},
  author={Huang, Cheng-Zhi Anna and Vaswani, Ashish and Uszkoreit, Jakob and Simon, Ian and Hawthorne, Curtis and Shazeer, Noam and Dai, Andrew M and Hoffman, Matthew D and Dinculescu, Monica and Eck, Douglas},
  booktitle={International Conference on Learning Representations},
  year={2019}
}

@inproceedings{zhou2019continuity,
  title={On the continuity of rotation representations in neural networks},
  author={Zhou, Yi and Barnes, Connelly and Lu, Jingwan and Yang, Jimei and Li, Hao},
  booktitle={Proceedings of the IEEE/CVF conference on computer vision and pattern recognition},
  pages={5745--5753},
  year={2019}
}

@inproceedings{chen2024taming,
  title={Taming diffusion probabilistic models for character control},
  author={Chen, Rui and Shi, Mingyi and Huang, Shaoli and Tan, Ping and Komura, Taku and Chen, Xuelin},
  booktitle={ACM SIGGRAPH 2024 Conference Papers},
  pages={1--10},
  year={2024}
}

@inproceedings{mao2017least,
  title={Least squares generative adversarial networks},
  author={Mao, Xudong and Li, Qing and Xie, Haoran and Lau, Raymond YK and Wang, Zhen and Paul Smolley, Stephen},
  booktitle={Proceedings of the IEEE international conference on computer vision},
  pages={2794--2802},
  year={2017}
}

@inproceedings{mescheder2018training,
  title={Which training methods for GANs do actually converge?},
  author={Mescheder, Lars and Geiger, Andreas and Nowozin, Sebastian},
  booktitle={International conference on machine learning},
  pages={3481--3490},
  year={2018},
  organization={PMLR}
}

@inproceedings{loshchilovdecoupled,
  title={Decoupled Weight Decay Regularization},
  author={Loshchilov, Ilya and Hutter, Frank},
  booktitle={International Conference on Learning Representations},
  year={2019}
}

@article{starke2022deepphase,
  title={Deepphase: Periodic autoencoders for learning motion phase manifolds},
  author={Starke, Sebastian and Mason, Ian and Komura, Taku},
  journal={ACM Transactions on Graphics (ToG)},
  volume={41},
  number={4},
  pages={1--13},
  year={2022},
  publisher={ACM New York, NY, USA}
}

@article{yi2021transpose,
  title={Transpose: Real-time 3d human translation and pose estimation with six inertial sensors},
  author={Yi, Xinyu and Zhou, Yuxiao and Xu, Feng},
  journal={ACM Transactions On Graphics (TOG)},
  volume={40},
  number={4},
  pages={1--13},
  year={2021},
  publisher={ACM New York, NY, USA}
}

@article{zhang2018mode,
  title={Mode-adaptive neural networks for quadruped motion control},
  author={Zhang, He and Starke, Sebastian and Komura, Taku and Saito, Jun},
  journal={ACM Transactions on Graphics (ToG)},
  volume={37},
  number={4},
  pages={1--11},
  year={2018},
  publisher={ACM New York, NY, USA}
}

@inproceedings{lee2023same,
  title={Same: Skeleton-agnostic motion embedding for character animation},
  author={Lee, Sunmin and Kang, Taeho and Park, Jungnam and Lee, Jehee and Won, Jungdam},
  booktitle={SIGGRAPH Asia 2023 Conference Papers},
  pages={1--11},
  year={2023}
}

@article{holden2017phase,
  title={Phase-functioned neural networks for character control},
  author={Holden, Daniel and Komura, Taku and Saito, Jun},
  journal={ACM Transactions on Graphics (TOG)},
  volume={36},
  number={4},
  pages={1--13},
  year={2017},
  publisher={ACM New York, NY, USA}
}

@article{10.1145/566654.566605,
author = {Kovar, Lucas and Gleicher, Michael and Pighin, Fr\'{e}d\'{e}ric},
title = {Motion graphs},
year = {2002},
issue_date = {July 2002},
publisher = {Association for Computing Machinery},
address = {New York, NY, USA},
volume = {21},
number = {3},
issn = {0730-0301},
url = {https://doi.org/10.1145/566654.566605},
doi = {10.1145/566654.566605},
journal = {ACM Trans. Graph.},
month = jul,
pages = {473–482},
numpages = {10}
}

@article{10.1145/1276377.1276510,
author = {Safonova, Alla and Hodgins, Jessica K.},
title = {Construction and optimal search of interpolated motion graphs},
year = {2007},
issue_date = {July 2007},
publisher = {Association for Computing Machinery},
address = {New York, NY, USA},
volume = {26},
number = {3},
issn = {0730-0301},
url = {https://doi.org/10.1145/1276377.1276510},
doi = {10.1145/1276377.1276510},
journal = {ACM Trans. Graph.},
month = jul,
pages = {106–es},
numpages = {12}
}

@article{peng2007interactive,
  title={Interactive and flexible motion transition},
  author={Peng, Jen-Yu and Lin, I-Chen and Chao, Jui-Hsiang and Chen, Yan-Ju and Juang, Gwo-Hao},
  journal={Computer Animation and Virtual Worlds},
  volume={18},
  number={4-5},
  pages={549--558},
  year={2007},
  publisher={Wiley Online Library}
}

@inproceedings{ren2010human,
  title={Human motion synthesis with optimization-based graphs},
  author={Ren, Cheng and Zhao, Liming and Safonova, Alla},
  booktitle={Computer Graphics Forum},
  volume={29},
  number={2},
  pages={545--554},
  year={2010},
  organization={Wiley Online Library}
}

@inproceedings{lee2002interactive,
  title={Interactive control of avatars animated with human motion data},
  author={Lee, Jehee and Chai, Jinxiang and Reitsma, Paul SA and Hodgins, Jessica K and Pollard, Nancy S},
  booktitle={Proceedings of the 29th annual conference on Computer graphics and interactive techniques},
  pages={491--500},
  year={2002}
}

@inproceedings{heck2007parametric,
  title={Parametric motion graphs},
  author={Heck, Rachel and Gleicher, Michael},
  booktitle={Proceedings of the 2007 symposium on Interactive 3D graphics and games},
  pages={129--136},
  year={2007}
}

@article{min2012motion,
  title={Motion graphs++ a compact generative model for semantic motion analysis and synthesis},
  author={Min, Jianyuan and Chai, Jinxiang},
  journal={ACM Transactions on Graphics (TOG)},
  volume={31},
  number={6},
  pages={1--12},
  year={2012},
  publisher={ACM New York, NY, USA}
}

@inproceedings{tao2023neural,
  title={Neural motion graph},
  author={Tao, Hongyu and Hou, Shuaiying and Zou, Changqing and Bao, Hujun and Xu, Weiwei},
  booktitle={SIGGRAPH Asia 2023 Conference Papers},
  pages={1--11},
  year={2023}
}

@inproceedings{guo2022tm2t,
  title={Tm2t: Stochastic and tokenized modeling for the reciprocal generation of 3d human motions and texts},
  author={Guo, Chuan and Zuo, Xinxin and Wang, Sen and Cheng, Li},
  booktitle={European Conference on Computer Vision},
  pages={580--597},
  year={2022},
  organization={Springer}
}

@article{jiang2023motiongpt,
  title={Motiongpt: Human motion as a foreign language},
  author={Jiang, Biao and Chen, Xin and Liu, Wen and Yu, Jingyi and Yu, Gang and Chen, Tao},
  journal={Advances in Neural Information Processing Systems},
  volume={36},
  pages={20067--20079},
  year={2023}
}

@inproceedings{zhang2023generating,
  title={Generating human motion from textual descriptions with discrete representations},
  author={Zhang, Jianrong and Zhang, Yangsong and Cun, Xiaodong and Zhang, Yong and Zhao, Hongwei and Lu, Hongtao and Shen, Xi and Shan, Ying},
  booktitle={Proceedings of the IEEE/CVF conference on computer vision and pattern recognition},
  pages={14730--14740},
  year={2023}
}

@inproceedings{pinyoanuntapong2024mmm,
  title={Mmm: Generative masked motion model},
  author={Pinyoanuntapong, Ekkasit and Wang, Pu and Lee, Minwoo and Chen, Chen},
  booktitle={Proceedings of the IEEE/CVF Conference on Computer Vision and Pattern Recognition},
  pages={1546--1555},
  year={2024}
}

@article{starke2024categorical,
  title={Categorical codebook matching for embodied character controllers},
  author={Starke, Sebastian and Starke, Paul and He, Nicky and Komura, Taku and Ye, Yuting},
  journal={ACM Transactions on Graphics (TOG)},
  volume={43},
  number={4},
  pages={1--14},
  year={2024},
  publisher={ACM New York, NY, USA}
}

@inproceedings{lu2025scamo,
  title={Scamo: Exploring the scaling law in autoregressive motion generation model},
  author={Lu, Shunlin and Wang, Jingbo and Lu, Zeyu and Chen, Ling-Hao and Dai, Wenxun and Dong, Junting and Dou, Zhiyang and Dai, Bo and Zhang, Ruimao},
  booktitle={Proceedings of the Computer Vision and Pattern Recognition Conference},
  pages={27872--27882},
  year={2025}
}

@inproceedings{li2024walkthedog,
  title={Walkthedog: Cross-morphology motion alignment via phase manifolds},
  author={Li, Peizhuo and Starke, Sebastian and Ye, Yuting and Sorkine-Hornung, Olga},
  booktitle={ACM SIGGRAPH 2024 Conference Papers},
  pages={1--10},
  year={2024}
}

@article{vaswani2017attention,
  title={Attention is all you need},
  author={Vaswani, Ashish and Shazeer, Noam and Parmar, Niki and Uszkoreit, Jakob and Jones, Llion and Gomez, Aidan N and Kaiser, {\L}ukasz and Polosukhin, Illia},
  journal={Advances in neural information processing systems},
  volume={30},
  year={2017}
}

@inproceedings{guo2022generating,
  title={Generating diverse and natural 3d human motions from text},
  author={Guo, Chuan and Zou, Shihao and Zuo, Xinxin and Wang, Sen and Ji, Wei and Li, Xingyu and Cheng, Li},
  booktitle={Proceedings of the IEEE/CVF conference on computer vision and pattern recognition},
  pages={5152--5161},
  year={2022}
}

@inproceedings{zheng2023online,
  title={Online clustered codebook},
  author={Zheng, Chuanxia and Vedaldi, Andrea},
  booktitle={Proceedings of the IEEE/CVF International Conference on Computer Vision},
  pages={22798--22807},
  year={2023}
}

@article{harvey2020robust,
  title={Robust motion in-betweening},
  author={Harvey, F{\'e}lix G and Yurick, Mike and Nowrouzezahrai, Derek and Pal, Christopher},
  journal={ACM Transactions on Graphics (TOG)},
  volume={39},
  number={4},
  pages={60--1},
  year={2020},
  publisher={ACM New York, NY, USA}
}

@article{tang2022real,
  title={Real-time controllable motion transition for characters},
  author={Tang, Xiangjun and Wang, He and Hu, Bo and Gong, Xu and Yi, Ruifan and Kou, Qilong and Jin, Xiaogang},
  journal={ACM Transactions on Graphics (TOG)},
  volume={41},
  number={4},
  pages={1--10},
  year={2022},
  publisher={ACM New York, NY, USA}
}

@inproceedings{Tang_2023, series={SIGGRAPH ’23},
   title={RSMT: Real-time Stylized Motion Transition for Characters},
   url={http://dx.doi.org/10.1145/3588432.3591514},
   DOI={10.1145/3588432.3591514},
   booktitle={Special Interest Group on Computer Graphics and Interactive Techniques Conference Conference Proceedings},
   publisher={ACM},
   author={Tang, Xiangjun and Wu, Linjun and Wang, He and Hu, Bo and Gong, Xu and Liao, Yuchen and Li, Songnan and Kou, Qilong and Jin, Xiaogang},
   year={2023},
   month=jul, pages={1–10},
   collection={SIGGRAPH ’23} }

@article{Kim_2022,
   title={Conditional motion in-betweening},
   volume={132},
   ISSN={0031-3203},
   url={http://dx.doi.org/10.1016/j.patcog.2022.108894},
   DOI={10.1016/j.patcog.2022.108894},
   journal={Pattern Recognition},
   publisher={Elsevier BV},
   author={Kim, Jihoon and Byun, Taehyun and Shin, Seungyoun and Won, Jungdam and Choi, Sungjoon},
   year={2022},
   month=dec, pages={108894} }

@article{qin2022motion,
  title={Motion In-Betweening via Two-Stage Transformers.},
  author={Qin, Jia and Zheng, Youyi and Zhou, Kun},
  journal={ACM Trans. Graph.},
  volume={41},
  number={6},
  pages={184--1},
  year={2022}
}

@article{oreshkin2023motion,
  title={Motion In-Betweening via Deep $\Delta$-Interpolator},
  author={Oreshkin, Boris N and Valkanas, Antonios and Harvey, F{\'e}lix G and M{\'e}nard, Louis-Simon and Bocquelet, Florent and Coates, Mark J},
  journal={IEEE Transactions on Visualization and Computer Graphics},
  volume={30},
  number={8},
  pages={5693--5704},
  year={2023},
  publisher={IEEE}
}

@inproceedings{hong2024long,
  title={Long-term Motion In-betweening via Keyframe Prediction},
  author={Hong, Seokhyeon and Kim, Haemin and Cho, Kyungmin and Noh, Junyong},
  booktitle={Computer Graphics Forum},
  volume={43},
  number={8},
  pages={e15171},
  year={2024},
  organization={Wiley Online Library}
}

@inproceedings{akhoundi2025silk,
  title={SILK: Smooth InterpoLation frameworK for motion in-betweening},
  author={Akhoundi, Elly and Ling, Hung Yu and Deshmukh, Anup Anand and B{\"u}tepage, Judith},
  booktitle={Proceedings of the Computer Vision and Pattern Recognition Conference},
  pages={2900--2909},
  year={2025}
}

@article{agrawal2024skel,
  title={SKEL-Betweener: a Neural Motion Rig for Interactive Motion Authoring},
  author={Agrawal, Dhruv and Buhmann, Jakob and Borer, Dominik and Sumner, Robert W and Guay, Martin},
  journal={ACM Transactions on Graphics (TOG)},
  volume={43},
  number={6},
  pages={1--11},
  year={2024},
  publisher={ACM New York, NY, USA}
}

@inproceedings{tevethuman,
  title={Human Motion Diffusion Model},
  author={Tevet, Guy and Raab, Sigal and Gordon, Brian and Shafir, Yoni and Cohen-or, Daniel and Bermano, Amit Haim},
  booktitle={The Eleventh International Conference on Learning Representations},
  year={2023}
}

@inproceedings{cohan2024flexible,
  title={Flexible motion in-betweening with diffusion models},
  author={Cohan, Setareh and Tevet, Guy and Reda, Daniele and Peng, Xue Bin and van de Panne, Michiel},
  booktitle={ACM SIGGRAPH 2024 Conference Papers},
  pages={1--9},
  year={2024}
}

@inproceedings{goel2025generative,
  title={Generative Motion Infilling from Imprecisely Timed Keyframes},
  author={Goel, Purvi and Zhang, Haotian and Liu, C Karen and Fatahalian, Kayvon},
  booktitle={Computer Graphics Forum},
  pages={e70060},
  year={2025},
  organization={Wiley Online Library}
}

@inproceedings{zhang2025motion,
  title={Motion In-Betweening for Densely Interacting Characters},
  author={Zhang, Xiaotang and Chang, Ziyi and Men, Qianhui and Shum, Hubert PH},
  booktitle={Proceedings of the SIGGRAPH Asia 2025 Conference Papers},
  pages={1--11},
  year={2025}
}

@article{he2022nemf,
  title={Nemf: Neural motion fields for kinematic animation},
  author={He, Chengan and Saito, Jun and Zachary, James and Rushmeier, Holly and Zhou, Yi},
  journal={Advances in Neural Information Processing Systems},
  volume={35},
  pages={4244--4256},
  year={2022}
}

@inproceedings{chu2024real,
  title={Real-time diverse motion in-betweening with space-time control},
  author={Chu, Yuchen and Yang, Zeshi},
  booktitle={Proceedings of the 17th ACM SIGGRAPH Conference on Motion, Interaction, and Games},
  pages={1--8},
  year={2024}
}

@inproceedings{bae2025less,
  title={Less is more: Improving motion diffusion models with sparse keyframes},
  author={Bae, Jinseok and Hwang, Inwoo and Lee, Young-Yoon and Guo, Ziyu and Liu, Joseph and Ben-Shabat, Yizhak and Kim, Young Min and Kapadia, Mubbasir},
  booktitle={Proceedings of the IEEE/CVF International Conference on Computer Vision},
  pages={11069--11078},
  year={2025}
}

@inproceedings{karunratanakul2023guided,
  title={Guided motion diffusion for controllable human motion synthesis},
  author={Karunratanakul, Korrawe and Preechakul, Konpat and Suwajanakorn, Supasorn and Tang, Siyu},
  booktitle={Proceedings of the IEEE/CVF international conference on computer vision},
  pages={2151--2162},
  year={2023}
}

@article{xie2023omnicontrol,
  title={Omnicontrol: Control any joint at any time for human motion generation},
  author={Xie, Yiming and Jampani, Varun and Zhong, Lei and Sun, Deqing and Jiang, Huaizu},
  journal={arXiv preprint arXiv:2310.08580},
  year={2023}
}

@inproceedings{dai2024motionlcm,
  title={Motionlcm: Real-time controllable motion generation via latent consistency model},
  author={Dai, Wenxun and Chen, Ling-Hao and Wang, Jingbo and Liu, Jinpeng and Dai, Bo and Tang, Yansong},
  booktitle={European Conference on Computer Vision},
  pages={390--408},
  year={2024},
  organization={Springer}
}

@inproceedings{pinyoanuntapong2025maskcontrol,
  title={Maskcontrol: Spatio-temporal control for masked motion synthesis},
  author={Pinyoanuntapong, Ekkasit and Saleem, Muhammad and Karunratanakul, Korrawe and Wang, Pu and Xue, Hongfei and Chen, Chen and Guo, Chuan and Cao, Junli and Ren, Jian and Tulyakov, Sergey},
  booktitle={Proceedings of the IEEE/CVF International Conference on Computer Vision},
  pages={9955--9965},
  year={2025}
}

@article{starke2023motion,
  title={Motion in-betweening with phase manifolds},
  author={Starke, Paul and Starke, Sebastian and Komura, Taku and Steinicke, Frank},
  journal={Proceedings of the ACM on Computer Graphics and Interactive Techniques},
  volume={6},
  number={3},
  pages={1--17},
  year={2023},
  publisher={ACM New York, NY, USA}
}

@inproceedings{bollo2018inertialization,
  title={Inertialization: High-performance animation transitions in’gears of war’},
  author={Bollo, David},
  booktitle={Proc. of GDC},
  volume={2016},
  year={2018}
}

@article{wang2008synthesis,
  title={Synthesis and evaluation of linear motion transitions},
  author={Wang, Jing and Bodenheimer, Bobby},
  journal={ACM Transactions on Graphics (TOG)},
  volume={27},
  number={1},
  pages={1--15},
  year={2008},
  publisher={ACM New York, NY, USA}
}

@article{li2008laziness,
  title={Laziness is a virtue: Motion stitching using effort minimization},
  author={Li, Lei and McCann, James and Faloutsos, Christos and Pollard, Nancy S},
  year={2008},
  publisher={Carnegie Mellon University}
}

@article{koyama2019precomputed,
  title={Precomputed optimal one-hop motion transition for responsive character animation},
  author={Koyama, Yuki and Goto, Masataka},
  journal={The Visual Computer},
  volume={35},
  number={6},
  pages={1131--1142},
  year={2019},
  publisher={Springer}
}

@inproceedings{kim2023recurrent,
  title={Recurrent motion refiner for locomotion stitching},
  author={Kim, Haemin and Cho, Kyungmin and Hong, Seokhyeon and Noh, Junyong},
  booktitle={Computer Graphics Forum},
  volume={42},
  number={6},
  pages={e14920},
  year={2023},
  organization={Wiley Online Library}
}

@article{starke2019neural,
  title={Neural state machine for character-scene interactions},
  author={Starke, Sebastian and Zhang, He and Komura, Taku and Saito, Jun},
  journal={ACM Transactions on Graphics},
  volume={38},
  number={6},
  pages={178},
  year={2019},
  publisher={ACM}
}

@misc{Mixamo,
    author = {Adobe},
    title = {Mixamo},
    year = {2025},
    howpublished = {\url{https://www.mixamo.com/}},
    note = {Accessed: 2025-01-16}
}

@inproceedings{guo2024momask,
  title={Momask: Generative masked modeling of 3d human motions},
  author={Guo, Chuan and Mu, Yuxuan and Javed, Muhammad Gohar and Wang, Sen and Cheng, Li},
  booktitle={Proceedings of the IEEE/CVF Conference on Computer Vision and Pattern Recognition},
  pages={1900--1910},
  year={2024}
}

@inproceedings{ghorbani2023zeroeggs,
  title={ZeroEGGS: Zero-shot Example-based Gesture Generation from Speech},
  author={Ghorbani, Saeed and Ferstl, Ylva and Holden, Daniel and Troje, Nikolaus F and Carbonneau, Marc-Andr{\'e}},
  booktitle={Computer Graphics Forum},
  volume={42},
  number={1},
  pages={206--216},
  year={2023},
  organization={Wiley Online Library}
}

@article{alexanderson2023listen,
  title={Listen, Denoise, Action! Audio-Driven Motion Synthesis with Diffusion Models},
  author={Alexanderson, Simon and Nagy, Rajmund and Beskow, Jonas and Henter, Gustav Eje},
  year={2023},
  issue_date={August 2023},
  publisher={ACM},
  volume={42},
  number={4},
  doi={10.1145/3592458},
  journal={ACM Trans. Graph.},
  articleno={44},
  numpages={20},
  pages={44:1--44:20}
}

@article{perez2021transflower,
  author={Valle-P{\'e}rez, Guillermo and Henter, Gustav Eje and Beskow, Jonas and Holzapfel, Andre and Oudeyer, Pierre-Yves and Alexanderson, Simon},
  title={Transflower: Probabilistic Autoregressive Dance Generation with Multimodal Attention},
  year={2021},
  issue_date={December 2021},
  publisher={ACM},
  volume={40},
  number={6},
  doi={10.1145/3478513.3480570},
  journal={ACM Trans. Graph.},
  articleno={195},
  numpages={14},
  pages={195:1--195:14}
}

@article{mason2022local,
    author = {Mason, Ian and Starke, Sebastian and Komura, Taku},
    title = {Real-Time Style Modelling of Human Locomotion via Feature-Wise Transformations and Local Motion Phases},
    year = {2022},
    publisher = {Association for Computing Machinery},
    address = {New York, NY, USA},
    volume = {5},
    number = {1},
    doi = {10.1145/3522618},
    journal = {Proceedings of the ACM on Computer Graphics and Interactive Techniques},
    month = {may},
    articleno = {6}
}

@article{huang2024interact,
      title={InterAct: Capture and Modelling of Realistic, Expressive and Interactive Activities between Two Persons in Daily Scenarios}, 
      author={Yinghao Huang and Leo Ho and Dafei Qin and Mingyi Shi and Taku Komura},
      year={2024},
      eprint={2405.11690},
      archivePrefix={arXiv},
      primaryClass={cs.CV}
}

\clearpage
\appendix

\section{Cluster Assignment}
After constructing clusters by training CVQ-VAE in Section 4.1 of the main paper, we assign a cluster to a new motion block by finding candidate codes from the codebook and choosing the final cluster among the candidates.
We adopt a candidate-based approach to avoid outliers, unlike the training process where the closest code is selected for the encoded vector.
When using only the nearest code, we observed that some samples exhibit root positions that differ significantly from the majority of the samples within the same cluster, and also deviate from the corresponding cluster clip.
Such outliers can produce false edge connections during graph construction, ultimately degrading motion quality.
We suspect this issue arises from the use of root-relative joint positions in $\mathbf{s}_t$, which may group motions with similar relative joint configurations but inconsistent root motion, leading to visual mismatches within a cluster.
Although the loss term $\mathcal{L}_{pos}$ mitigates this problem by incorporating world-frame positions, it does not fully ensure consistent root positions for all samples assigned to the same cluster.

To address this, we first select candidate codes and then compare the root position to identify the final cluster ID.
We encode $\mathbf{S}_A$ into $\mathbf{z}_A$, and retrieve the $n$ closest cluster codes $\hat{\mathbf{c}}_{a_1}, \hat{\mathbf{c}}_{a_2}, ..., \hat{\mathbf{c}}_{a_n}$ in terms of cosine distance, along with their cluster clips, $\hat{\mathbf{S}}_{a_1}, \hat{\mathbf{S}}_{a_2}, \dots, \hat{\mathbf{S}}_{a_n}$.
The final cluster ID $k_A$ is obtained by selecting the candidate cluster whose cluster clip has the minimum Euclidean distance in root position from the input motion $\mathbf{S}_A$:
\[
k_A = \underset{k \in \{a_1, a_2, \dots, a_n\}}{\arg\min} \;
\frac{1}{T_l} \sum_{i=1}^{T_l}\|\mathbf{p}_{A, i} - \hat{\mathbf{p}}_{k, i}\|_2.
\]
\noindent
$\mathbf{p}_{A, i}$ denotes the root position at frame $i$ in $\mathbf{S}_A$, and $\hat{\mathbf{p}}_{k, i}$ denotes the same value from $\hat{\mathbf{S}}_k$.
We mapped all motion blocks in the dataset to clusters and selected the minimum number of candidates that avoided outlier assignment, which was $n=3$.

\section{Directly Connected Paths}\label{appen:direct}
During the input processing described in Section 6.1.2 of the main paper, we construct guides differently when the start and goal nodes are directly connected in the cluster transition graph, such that the resulting path consists only of these two nodes.
Instead of using cluster clips, we insert a single block of $T_l$ frames between $\mathbf{S}_A$ and $\mathbf{S}_B$, which is the minimum transition length in our method.
Then these frames are filled by interpolating each feature in $\mathbf{S}_A$ and $\mathbf{S}_B$. 
In this case, the resulting guide $\mathbf{S}_g$ consists of $N_b+1$ motion blocks.

\section{Implementation Details}
\subsubsection{CVQ-VAE training details}
We trained CVQ-VAE on a motion block dataset created from training sequences in LAFAN1.
The model was trained for 100 epochs with a batch size of 64, which required about 2.2 hours.
To select anchors for online reinitialization, probabilistic random vectors~\cite{zheng2023online} were employed.
We used an AdamW optimizer~\cite{loshchilovdecoupled} with a learning rate of $10^{-4}$ on an Nvidia GeForce RTX 3090 GPU. 
The learning rate was gradually reduced using cosine function, reaching a minimum value of $10^{-5}$ at the last epoch.

% section 5.2
\subsubsection{Cluster pathfinding weight tuning}
Adjusting the weights $\lambda_{code}$ and $\lambda_{bnd}$ in Equation~1, and $\lambda_h$, $\lambda_{q}$, $\lambda_{\Delta{p}}$, and $\lambda_{\Delta{q}}$ in Equation~3 of Section~5.2 of the main paper strongly influences the resulting path $\mathbf{m}$ and the quality of the generated motion.
To facilitate tuning, we precomputed all pairwise distances between valid codebook entries, both for cluster codes and for cluster clips.
We then normalized the range of all precomputed distance values to $[0, 1]$.
The distance values for each feature were divided by the feature-wise mean value to reduce the scale imbalance between the terms.
The weights were then tuned by evaluating motion quality in the subsequent motion generation stage.

\subsubsection{Motion generator training details}
The motion generator was trained on subsequences of LAFAN1 with varying lengths. 
It was trained for 100 epochs with a batch size of 64, which required about 25 hours.
We adopted the AdamW optimizer on an Nvidia GeForce RTX 3090 GPU.
The learning rate was increased to $10^{-4}$ during a linear warm-up stage for the first 1\% of the total iteration.
After warm-up, cosine annealing was applied to gradually decrease the learning rate to zero.

The adversarial training started only after the warm-up stage to prevent the discriminator from dominating the training process.
For discriminator training, we used an Adam optimizer with a learning rate of $10^{-5}$, which was slower than the motion generator.
The learning rate was also annealed following cosine function, reaching $10^{-6}$ at the last epoch.

\begin{figure}[b]
    \centering
    \includegraphics[width=0.47\textwidth]{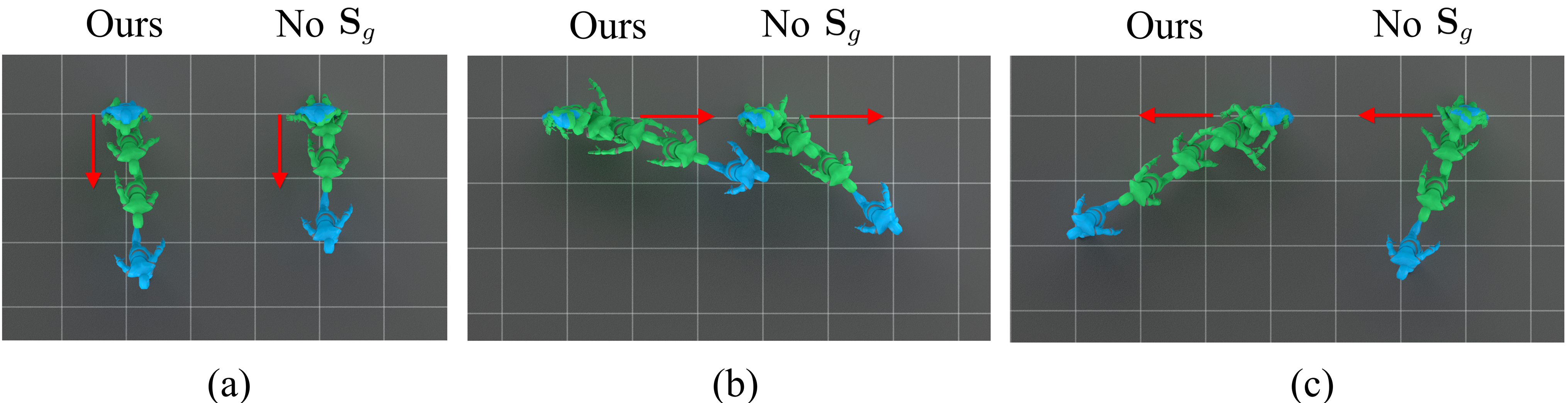}
    \caption{Comparison between our method and the model trained without $\mathbf{S}_g$. Results are visualized every 16 frames from the top view and the sequence from No $\mathbf{S}_g$ is translated for visibility. The red arrows indicate the input control vector.}
    \label{fig:control_noguide}
\end{figure}

\section{Effect of Guide for Direction Control}
In addition to the results reported in Section~7.4.1 of the main paper, the guide input $\mathbf{S}_g$ improves the model's responsiveness to control signals in the direction control scenario described in Section~7.3
Given the same inputs, we provided the same control vectors to each model and compared the resulting transition trajectories as shown in Figure~\ref{fig:control_noguide}.
While both models performed similarly under a forward control input, differences became apparent when significant directional changes are required.
In these cases, the motion generated by our method adhered more closely to the control input, exhibiting sharper body turns compared to the model trained without $\mathbf{S}_g$.
This suggests that $\mathbf{S}_g$ provides additional cues on what poses to generate, complementing the information provided by the second input motion.
Moreover, as the guide directly conditions the generation process, transition generation can be easily controlled by adjusting the pathfinding step.

\section{Additional Comparison}

\begin{figure}
    \centering
    \includegraphics[width=0.3\textwidth]{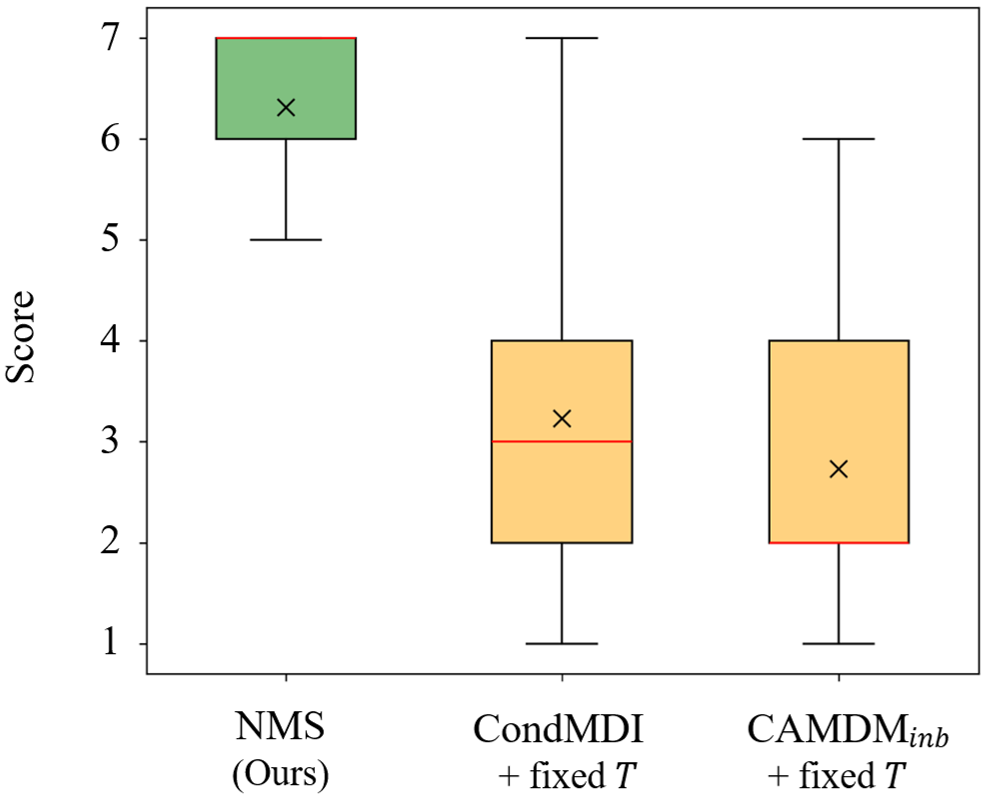}
    \caption{\hm{User study results shown as box plots.}}
    \label{fig:userstudy}
\end{figure}

We additionally compare our method with another diffusion-based model that generates future motion from past motion~\cite{chen2024taming}.
Because their original model was not designed for motion in-betweening, we adapted it to the in-betweening setting and refer to the modified model as $\mathrm{CAMDM}_{inb}$.
Specifically, we used the past motion token from the original formulation and input second motion along with noisy frames, allowing the network to observe the target motion.
The adapted model was trained on variable-length motion sequences from the LAFAN1 training set, as with the other baselines.

We evaluate $\mathrm{CAMDM}_{inb}$, CondMDI, and NMS with the same samples and metrics introduced in Section 7.6 of the main paper.
The qualitative comparison is provided in Figure~\ref{fig:compare_qual}. 
Similar to CondMDI, $\mathrm{CAMDM}_{inb}$ with a fixed total length $T$ produced overly smoothed motion and boundary discontinuities.
Combining the model with BC did not improve the performance, as shown in the second-to-last row of the figure.
The generated motion lacked detail and struggled to match the second motion in Sample (a), and noticeable discontinuities remained at the boundaries in Sample (b).

These observations are consistent with our quantitative evaluation.
As presented in Table~\ref{tab:metric}, $\mathrm{CAMDM}_{inb}$ exhibited high jitter, despite the moderate performance on FID and detail distance. 
We assume that this is because the generated motions often deviate from the target motions, especially when the given input is significantly different from training samples as in Mixamo + 100Style.
In contrast, the metric scores demonstrate that ours successfully connects two given motions with minimal jitter, while maintaining low FID and detail distance scores across datasets.

We further report user study results in Figure~\ref{fig:userstudy} to compare the visual quality of the results produced by our method relative to fixed-length $\mathrm{CAMDM}_{inb}$.
Our method achieved the highest score, while $\mathrm{CAMDM}_{inb}$ obtained an average score similar to CondMDI.
These results indicates that our method provides better overall visual quality than $\mathrm{CAMDM}_{inb}$.

\begin{figure}
    \centering
    \includegraphics[width=0.45\textwidth]{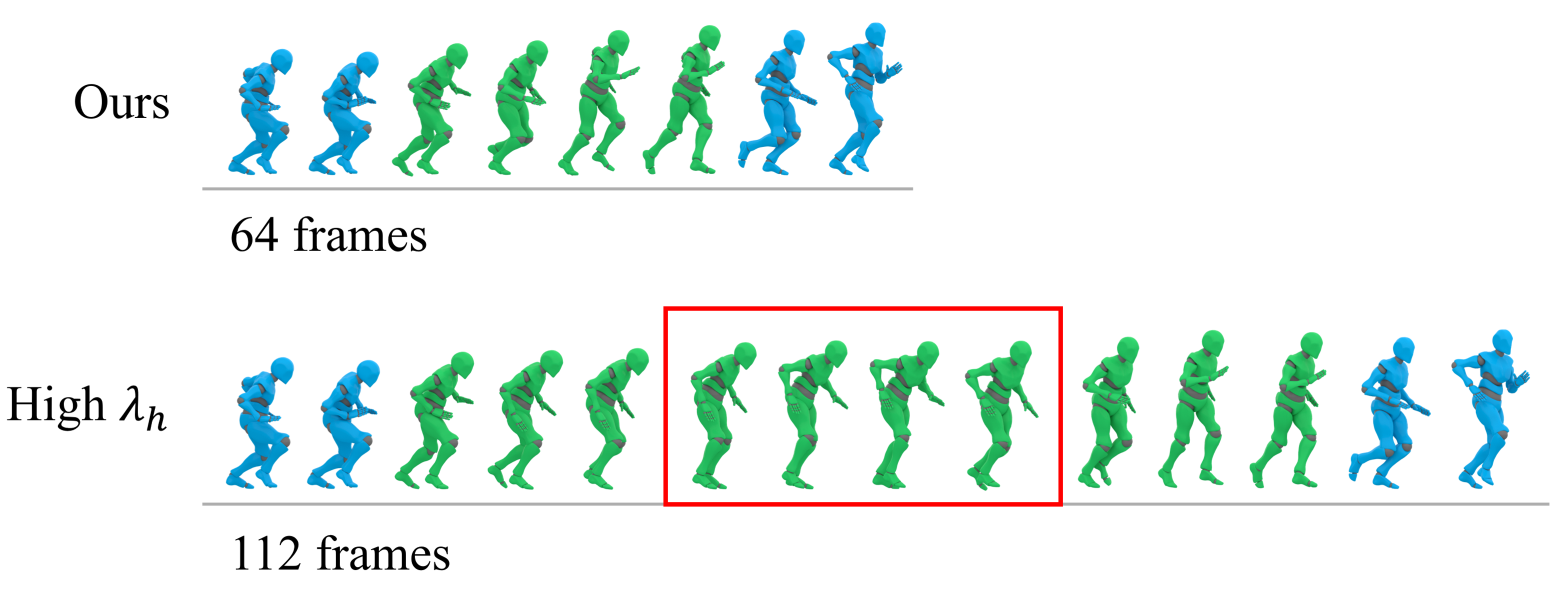}
    \caption{Comparison of our method using the default value of $\lambda_h$ and twice the default value. The red box highlights the difference between the two results.}
    \label{fig:hyper}
\end{figure}

\begin{figure*}
    \centering
    \includegraphics[width=0.8\textwidth]{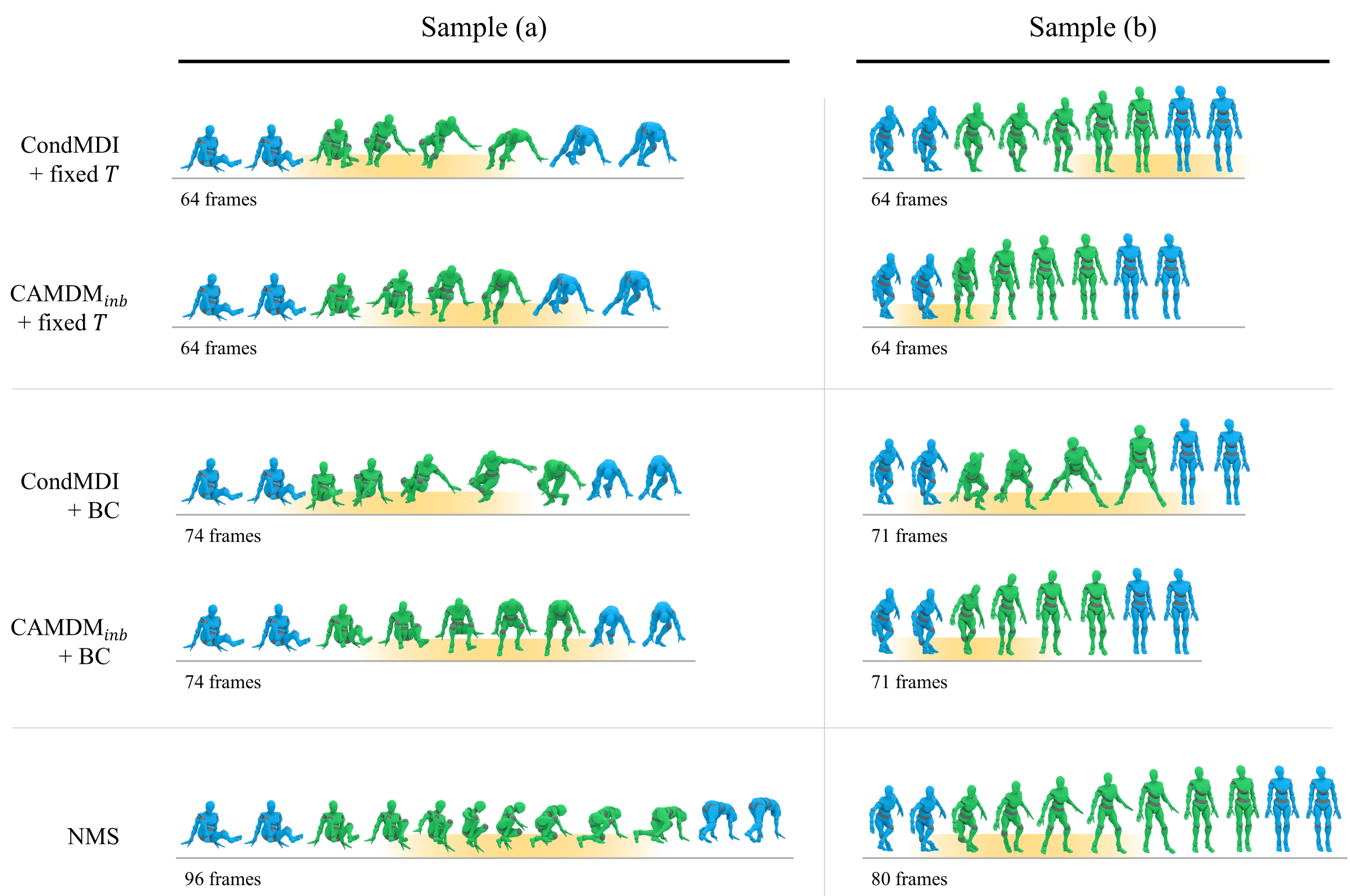}
    \caption{Qualitative comparison with other methods. In this figure, green characters indicate the poses edited or generated by each method. Notable poses in each results are highlighted in yellow.}
    \label{fig:compare_qual}
\end{figure*}

\section{Hyperparameter Ablation}
We evaluated the sensitivity of the cluster pathfinding cost weights introduced in Section 5.2, namely $\lambda_{code}$, $\lambda_{h}$, $\lambda_{q}$, $\lambda_{\Delta p}$, and $\lambda_{\Delta q}$.
We randomly sampled 1,000 input motion pairs from the $\mathrm{LAFAN1}_{within}$ dataset, fixed $\lambda_{bnd}=1$, and independently halved and doubled each weight while keeping the remaining weights fixed.
As shown in the Metrics columns of Table~\ref{tab:hyper}, although some alternative settings achieve slightly better scores on individual metrics, no single configuration consistently outperforms the default setting across all evaluation metrics within the tested range.

Changing the cost weights had impact on selecting cluster paths and the resulting motion lengths.
As reflected in the average transition lengths reported in the last column of Table~\ref{tab:hyper}, different weight configurations produced motions of different lengths even for the same input pair.
We selected our setting over High $\lambda_{h}$ because ours achieved the best detail distance score, while High $\lambda_{h}$ often resulted in lingering motion before it reached the second input motion.
This qualitative difference is illustrated in Figure~\ref{fig:hyper}, where ours generates a motion with 64 frames, whereas High $\lambda_h$ produces a longer motion with 112 frames, going through additional footsteps before the second input motion.
An interesting direction for future work would be to automatically learn these weights, adapting them to the input motions or user preferences.

\begin{table*}
\centering
\small
\caption{\hm{Comparison of FID, detail distance, jitter, FS, and GP. Best scores are shown in bold, and second-best scores are underlined.}}
\label{tab:metric}
\hm{
\begin{tabularx}{0.99\textwidth}{
l
C{0.6cm} C{0.6cm} C{0.8cm} C{0.6cm} C{0.6cm}
C{0.6cm} C{0.6cm} C{0.8cm} C{0.6cm} C{0.6cm}
C{0.6cm} C{0.6cm} C{0.8cm} C{0.6cm} C{0.6cm}
}
\toprule
& \multicolumn{5}{c}{$\mathrm{LAFAN1}_{within}$}
& \multicolumn{5}{c}{$\mathrm{LAFAN1}_{random}$}
& \multicolumn{5}{c}{Mixamo + 100Style} \\
\cmidrule(lr){2-6}
\cmidrule(lr){7-11}
\cmidrule(lr){12-16}
Method
& \begin{tabular}[c]{@{}c@{}}FID$\downarrow$\end{tabular} 
& \begin{tabular}[c]{@{}c@{}}Detail\\dist.$\downarrow$\end{tabular} 
& \begin{tabular}[c]{@{}c@{}}Jitter$\downarrow$\end{tabular}
& \begin{tabular}[c]{@{}c@{}}FS$\downarrow$\end{tabular}
& \begin{tabular}[c]{@{}c@{}}GP$\downarrow$\end{tabular}
& \begin{tabular}[c]{@{}c@{}}FID$\downarrow$\end{tabular} 
& \begin{tabular}[c]{@{}c@{}}Detail\\dist.$\downarrow$\end{tabular} 
& \begin{tabular}[c]{@{}c@{}}Jitter$\downarrow$\end{tabular}
& \begin{tabular}[c]{@{}c@{}}FS$\downarrow$\end{tabular}
& \begin{tabular}[c]{@{}c@{}}GP$\downarrow$\end{tabular}
& \begin{tabular}[c]{@{}c@{}}FID$\downarrow$\end{tabular} 
& \begin{tabular}[c]{@{}c@{}}Detail\\dist.$\downarrow$\end{tabular} 
& \begin{tabular}[c]{@{}c@{}}Jitter$\downarrow$\end{tabular}
& \begin{tabular}[c]{@{}c@{}}FS$\downarrow$\end{tabular}
& \begin{tabular}[c]{@{}c@{}}GP$\downarrow$\end{tabular} \\
\midrule
CondMDI + fixed $T$ & \underline{0.58} & \underline{7.29} & 3797.48 & 0.54 & \textbf{0.09} & 8.05 & \underline{15.90} & 4510.02 & 1.14 & \textbf{0.09} & 14.58 & 31.56 & 5131.86 & 1.46 & \underline{0.06} \\
$\mathrm{CAMDM}_{inb}$ + fixed $T$ & 0.91 & 10.42 & \underline{3385.96} & 0.47 & \textbf{0.09} & \underline{7.80} & 20.51 & \underline{3712.66} & 0.98 & 0.11 & \textbf{13.40} & \underline{29.89} & \underline{4008.95} & 1.22 & \underline{0.06}\\
\midrule
CondMDI + BC & 4.03 & 32.22 & 10656.29 & 1.88 & 0.14 & 20.61 & 58.78 & 17456.14 & 3.53 & 0.22 & 30.33 & 53.08 & 19912.32 & 4.12 & 0.16 \\
$\mathrm{CAMDM}_{inb}$ + BC & 1.28 & 13.18 & 3985.72 & \underline{0.39} & \underline{0.10} & 11.67 & 21.07 & 6050.19 & \underline{0.64} & 0.19 & 15.97 & 30.78 & 6925.71 & \underline{0.68} & 0.11 \\
\midrule
NMS (Ours) & \textbf{0.20} & \textbf{5.18} & \textbf{382.89} & \textbf{0.23} & \textbf{0.09} & \textbf{7.60} & \textbf{15.86} & \textbf{385.00} & \textbf{0.38} & \underline{0.10} & \underline{14.26} & \textbf{26.54} & \textbf{589.82} & \textbf{0.52} & \textbf{0.05} \\
\bottomrule
\end{tabularx}
}
\end{table*}

\begin{table*}
\centering
\small
\setlength{\tabcolsep}{4.2pt}
\renewcommand{\arraystretch}{1.12}
\caption{Comparison of FID, detail distance, jitter, foot sliding, and ground penetration across different hyperparameter settings. The first row reports the results obtained with our default setting. The remaining rows show the results obtained by halving or doubling one parameter at a time. Foot sliding is scaled by $10^{3}$ and ground penetration is represented in centimeters. Best scores are indicated in bold, and second-best scores are underlined.}
\label{tab:hyper}
{
\begin{tabular}{
l
*{5}{C{0.5cm}}
*{5}{C{0.9cm}}
C{1.2cm}
}
\toprule
& \multicolumn{5}{c}{Parameter setting}
& \multicolumn{5}{c}{Metrics} 
& \multicolumn{1}{c}{Length} \\
\cmidrule(lr){2-6}
\cmidrule(lr){7-11}
\cmidrule(lr){12-12}

Variation
& \begin{tabular}[c]{@{}c@{}}$\lambda_{code}$\end{tabular} 
& \begin{tabular}[c]{@{}c@{}}$\lambda_{h}$\end{tabular} 
& \begin{tabular}[c]{@{}c@{}}$\lambda_{q}$\end{tabular}
& \begin{tabular}[c]{@{}c@{}}$\lambda_{\Delta{p}}$\end{tabular}
& \begin{tabular}[c]{@{}c@{}}$\lambda_{\Delta{q}}$\end{tabular}
& \begin{tabular}[c]{@{}c@{}}FID$\downarrow$\end{tabular} 
& \begin{tabular}[c]{@{}c@{}}Detail\\dist.$\downarrow$\end{tabular} 
& \begin{tabular}[c]{@{}c@{}}Jitter$\downarrow$\end{tabular}
& \begin{tabular}[c]{@{}c@{}}Foot\\sliding$\downarrow$\end{tabular}
& \begin{tabular}[c]{@{}c@{}}Ground\\pene.$\downarrow$\end{tabular}
& \begin{tabular}[c]{@{}c@{}}Avg.\\length\end{tabular} \\
\midrule
Ours & 0.2 & 1.0 & 1.0 & 0.5 & 0.5 & 0.620 & \textbf{5.616} & 378.083 & 0.253 & \underline{0.093} & 58.416\\
\midrule
Low $\lambda_{code}$ & 0.1 & 1.0 & 1.0 & 0.5 & 0.5 & 0.619 & 5.699 & 377.089 & 0.253 & \underline{0.093} & 58.704 \\
High $\lambda_{code}$ & 0.4 & 1.0 & 1.0 & 0.5 & 0.5 & 0.622 & \underline{5.641} & 377.671 & 0.251 & \underline{0.093} & 58.432 \\
\midrule
Low $\lambda_{h}$ & 0.2 & 0.5 & 1.0 & 0.5 & 0.5 & 0.620 & 5.704 & 377.746 & \textbf{0.238} & \underline{0.093} & 58.416 \\
High $\lambda_{h}$ & 0.2 & 2.0 & 1.0 & 0.5 & 0.5 & \textbf{0.616} & 5.847 & \textbf{375.124} & 0.244 & \underline{0.093} & 59.584\\
\midrule
Low $\lambda_{q}$ & 0.2 & 1.0 & 0.5 & 0.5 & 0.5 & 0.621 & 5.678 & 377.085 & \underline{0.241} & \textbf{0.092} & 58.736\\
High $\lambda_{q}$ & 0.2 & 1.0 & 2.0 & 0.5 & 0.5 & 0.620 & 5.732 & 377.466 & 0.245 & \underline{0.093} & 58.928\\
\midrule
Low $\lambda_{\Delta{p}}$ & 0.2 & 1.0 & 1.0 & 0.25 & 0.5 & 0.620 & 5.696 & 377.358 & 0.253 & \underline{0.093} & 58.784\\
High $\lambda_{\Delta{p}}$ & 0.2 & 1.0 & 1.0 & 1.0 & 0.5 & 0.622 & 5.677 & 377.035 & 0.249 & \underline{0.093} & 58.368\\
\midrule
Low $\lambda_{\Delta{q}}$ & 0.2 & 1.0 & 1.0 & 0.5 & 0.25 & 0.621 & 5.709 & 376.583 & 0.244 & \underline{0.093} & 58.816\\
High $\lambda_{\Delta{q}}$ & 0.2 & 1.0 & 1.0 & 0.5 & 1.0 & \underline{0.618} & 5.663 & \underline{376.215} & 0.250 & \textbf{0.092} & 58.816\\
\bottomrule
\end{tabular}
}
\end{table*}

\end{document}